\documentclass{aastex}
\usepackage{emulateapj5}

\received{2002 August 11}
\begin{document}

\def\'#1{\ifx#1i{\accent"13\i}\else{\accent"13#1}\fi}

\def\BP{Ballesteros-Paredes}
\def\gamef{\gamma_{\rm e}}
\def\Jt{J_{\rm t}}
\def\Ms{M_{\rm s}}
\def\Pext{P_{\rm ext}}
\def\Pint{P_{\rm int}}
\def\nm{\bar n}
\def\nres{n_{\rm res}}
\def\np{n_{\rm peak}}
\def\nt{n_{\rm t}}
\def\tad{\tau_{\rm AD}}
\def\tfc{t_{\rm fc}}
\def\tfg{t_{\rm fg}}
\def\ts{t_{\rm s}}
\def\VS{V\'azquez-Semadeni}

\slugcomment{Submitted to The Astrophysical Journal}

\shorttitle{Lifetimes and Evolution of Molecular Cloud Cores}
\shortauthors{V\'azquez-Semadeni, Kim, Shadmehri \& Ballesteros-Paredes}

\title{The Lifetimes and Evolution of Molecular Cloud Cores}

\author{Enrique V\'azquez-Semadeni$^1$, Jongsoo Kim$^2$, Mohsen
Shadmehri$^3$ and Javier \BP$^1$} 
\affil{$^1$Centro de Radioastronom\'ia y Astrof\'isica, UNAM,
Apdo. Postal 72-3 (Xangari), Morelia, Michoac\'an 58089, M\'exico}
\email{e.vazquez, j.ballesteros@astrosmo.unam.mx}
\affil{$^2$ Korea Astronomy Observatory, San 61-1, Hwaam-Dong,
Yusong-Ku, Taejon 305-764, Korea}
\email{jskim@kao.re.kr}
\affil{$^3$Department of Physics, School of Science, Ferdowsi
University, Mashhad, Iran} \email{mshadmehri@science1.um.ac.ir}

\begin{abstract}
We discuss the lifetimes and evolution of clumps and cores formed
as turbulent density fluctuations in nearly isothermal molecular
clouds. In order to maintain a broad perspective, we consider both the
magnetic and non-magnetic cases. In the latter, we argue that clumps are
unlikely to reach a hydrostatic state if molecular clouds can in general
be described as single-phase media with an effective polytropic exponent
$\gamef < 4/3$. In this
case, clumps are expected to be short-lived,
either proceeding directly to collapse,
or else ``rebounding'' towards the mean pressure and density as the
parent cloud. Rebounding clumps are delayed in their re-expansion by
their self-gravity. From a simple Virial Theorem calculation, we find
re-expansion times $\sim 1$ Myr, i.e., of the order of a few local free-fall
times $\tfc \equiv L_{\rm J,c}/c$, where $L_{\rm J,c}=\sqrt{\pi c^2/G
\rho_{\rm c}}$ is the Jeans length at the clump's mean density
$\rho_{\rm c}$ and $c$ is the isothermal sound speed. 

In the magnetic case, we present a series of driven-turbulence,
ideal-MHD numerical
simulations in which we follow the evolution of clumps and cores in
relation to the magnetic criticality of their ``parent clouds'' (the
numerical boxes). In subcritical boxes, magnetostatic clumps do not
form. A minority of moderately-gravitationally bound clumps form 
which however are dispersed by the turbulence in $\lesssim 1.3$ Myr. An
estimate of the ambipolar diffusion (AD) time scale at the physical
conditions of these cores gives characteristic times $\gtrsim 1.3$ Myr,
suggesting that these few longer-lived cores can marginally be
``captured'' by AD 
to increase their mass-to-flux ratio and eventually collapse, although
on time scales not significantly longer than the dynamical ones. In
supercritical boxes, some cores manage to become 
locally supercritical and collapse in typical time scales of 2
$\tfc$ ($\sim$ 1 Myr). In the most supercritical simulation, a few
longer-lived cores are 
observed, which last for up to $\sim 3$ Myr, but these end up
re-expanding rather than collapsing, because they are sub-Jeans in spite
of being super-critical. 
Fewer clumps and cores form in these simulations than in their
non-magnetic counterpart.

Our results suggest that {\it a)} Not all cores observed 
in molecular clouds will necessarily form stars, and that a
class of ``failed cores'' should exist, which will eventually
re-disperse, and which may be related to the observed starless
cores. {\it b)} Cores may be
out-of-equilibrium, transient structures, rather than
quasi-magnetostatic configurations. {\it c)} The magnetic field may help
reduce the star formation efficiency by reducing the probability of core
formation, rather than by significantly delaying the collapse of
individual cores. 

\end{abstract}


\section{Introduction}\label{sec:intro}

One of the most important goals in the study of star formation is to
understand the state and physical conditions of the molecular cloud
cores from which the stars form. The prevailing view (which we
hereafter refer to as the ``standard (magnetic support) model'' of
star formation; e.g., Mouschovias 1976a,b; Shu, Adams \& Lizano 
1987) is that low-mass star-forming clumps evolve into dense cores along
a sequence of quasi-magnetostatic states as magnetic support is lost
by ambipolar diffusion. This process has a relatively long
characteristic time scale, typically estimated to be significantly
larger than the local clump free-fall time (e.g., Shu et al.\ 1987;
McKee et al.\ 1993). During this time, the contracting clump is
supported against its self-gravity by the magnetic field in the direction
perpendicular to it, and by a combination of thermal and
micro-turbulent pressures along the field (e.g., Lizano \& Shu 1989;
McLaughlin \& Pudritz 1996). 

The support from magnetic fields is included through the
consideration of the mass-to-magnetic flux ratio of the core. For
isolated or periodic structures, in which mass cannot be accreted from
the surroundings, and under ideal MHD conditions, the magnetic flux is
conserved, and therefore so is the mass-to-flux ratio (Chandrasekhar \& Fermi
1953; Mestel \& Spitzer 1956). Thus, self-gravity cannot overcome the
magnetic support 
if the mass-to-flux ratio is smaller than some critical value, and
collapse can only occur as the neutrals in the gas slowly slip across
field lines 
by ambipolar diffusion (see, e.g., Mestel \& Spitzer 1956; Mouschovias \&
Spitzer 1976; see also the reviews by Shu et al.\ 1987 and McKee
et al.\ 1993).  

Additionally, although in the 
standard model the magnetic support is fundamental, discussions of
non-magnetic equilibrium configurations are very often encountered in
the literature. These configurations have generally either finite or
diverging central densities. The former are generally based on the so-called
Bonnor-Ebert (BE; Ebert 1955; Bonnor 1956) spheres, which are 
solutions of the Lane-Emden equation, and can be either
truncated at some finite radius, beyond which a hotter, less dense
medium is assumed to confine the sphere, or infinite
(``extended''). Truncated configurations have recently been 
taken as ``templates'' for comparison with observations of column
density profiles in real cores (e.g., Johnstone et al.\ 2000; Alves,
Lada \& Lada 2001; Evans et al.\ 2001; Harvey et al.\ 2001), even
though Shu et al.\ (1987, sec. 4.1)
had already pointed out that such truncated structures may have
little relevance to actual cores in molecular clouds because they are
in contact with a large reservoir of cold gas, and that a maximum
contrast of 14 contradicts the observational evidence that
molecular clouds typically have H$_2$ densities ranging from $10^2$
cm$^{-3}$ (average) to well in excess of $10^6$ cm$^{-3}$ in cores (see,
e.g. Goldsmith 1987.) 
Various geometries other than spherical have also been considered for
non-magnetic equilibrium solutions  (e.g., Curry 2000; Lombardi \&
Bertin 2001). Concerning structures with diverging central densities,
the best known example is the singular isothermal sphere (e.g., Shu 1977).

On the other hand, it is well established that the molecular clouds
within which the cores form are turbulent, with linewidths that are
supersonic for scales $\gtrsim 0.05$ pc (e.g., Zuckerman \& Evans 1974;
Larson 1981; see also the review by Blitz 1993 and references therein),
generally interpreted as magnetohydrodynamic (MHD) supersonic turbulent
motions. 
Being supersonic (and possibly super-Alfv\'enic; see below), these
motions in general dominate the support 
against the clouds' self-gravity, with thermal pressure being dominant
only at subsonic scales (Padoan 1995; \VS, \BP\ \& Klessen 2003a). In this
environment, the molecular cloud clumps and cores are in fact likely to
be the turbulent density fluctuations themselves within 
the clouds (von Weizs\"acker 1951; Bania \& Lyon 1980; Scalo 1987;
Elmegreen 1993; \BP, \VS\ \& Scalo 1999;Klessen, Heitsch \& Mac Low
2000;  Padoan et al.\ 2001; Heitsch, Mac Low \& Klessen 2001). 

Moreover, in a turbulent environment, the density and magnetic field
strength are only very poorly correlated locally, at least when
gravity is not important in the formation of the density sturctures
(Passot, \VS\ \& Pouquet 1995; Padoan \& Nordlund 1999; Ostriker,
Stone \& Gammie 2001; Kim, Balsara \& Mac Low 2001; Passot \& \VS\
2003). Thus, the clumps and cores can in principle be formed with a {\it
distribution} of mass-to-flux ratios. For ideal isolated or periodic
configurations, this distribution is bounded
from above by the mass-to-flux ratio of the parent cloud (cf. \S
\ref{sec:magn_quali}) for as long
as ambipolar diffusion is negligible (i.e., at densities $\lesssim
10^4$ cm$^{-3}$). Real molecular clouds are not isolated nor periodic,
and can accrete mass along the magnetic field to eventually become
supercritical (Hartmann, Ballesteros-Paredes \& Bergin 2001), although
this is likely to occur on time scales much longer than those of core
evolution. So, for the purpose of studying core evolution, the
mass-to-flux ratio of the parent cloud can be considered constant.

Within this framework, if the parent cloud is subcritical, then in principle
supercritical cores can only form through the action of ambipolar
diffusion, in agreement 
with the scenario of the standard model. Indeed, numerical simulations
of supersonic turbulence in subcritical boxes without ambipolar
diffusion have shown that collapse does not occur (Ostriker, Gammie \&
Stone 1999; Heitsch et al.\ 2001).
On the other hand, if the parent cloud is supercritical, then in 
principle both sub- and supercritical substructures can form, even under
ideal MHD. The latter can
collapse if they are also gravitationally unstable with respect to the
thermal pressure, a case we will sometimes refer to as ``super-Jeans''. 

Observationally, several recent studies (e.g., Crutcher 1999, 2004;
Bourke et al.\ 2001; Crutcher, Heiles \& Troland 2002) suggest that
molecular clouds are the objects with highest mass-to-flux ratios
(within the uncertainties) in the hierarchy going from diffuse medium
to molecular cloud cores. Furthermore, Nakano (1998) has
suggested that the large observed column densities of cores are
difficult to reconcile with the relatively low column density
enhancements expected for subcritical cores with respect to their
parent clouds, while Padoan and coworkers (e.g., Padoan \& Nordlund
1999; Padoan at al.\ 1999; Padoan, Goodman \& Juvela 2003) have
performed several synthetic observations of cores in simulations of
MHD turbulence suggesting that molecular clouds are
super-Alfv\'enic. Finally, like their parent molecular clouds, the
clumps and cores (density peaks) within them are not isolated, but are
connected to their surrondings, and so their masses are not fixed,
especially in a turbulent environment. The non-isolated nature of the
cores has been taken into account within the standard model (e.g.,
Lizano \& Shu 1989; Curry 2000), but still under the assumption of a
quasi-static evolution.

There thus appears to exist a conceptual gap between the standard assumption
of quasi-hydrostatic clump/core evolution and the turbulent nature
of the clouds in which they form, the natural question being whether
cores formed dynamically in a turbulent medium can settle into the
(magneto-)hydrostatic equilibria
assumed in the standard model. Instead, they might as well be either
pushed directly into collapse, or re-expand and merge back with their
environment. Recently, Li \& Nakamura (2004) have addressed this issue
by considering the formation of dense cores in
two-dimensional simulations of sheet-like, strongly magnetized clouds
including a prescription for ambipolar diffusion.
In the present paper we consider the problem from a more general point
of view, by discussing clump and core formation in both strongly and
weakly (including non-magnetic) magnetized clouds, investigating
the lifetimes of clumps formed by transient turbulent fluctuations and
the evolution of their Jeans number and magnetic criticality in relation
to that of their parent molecular clouds. We begin (\S 
\ref{sec:assumptions}) by presenting and 
discussing the main assumptions that underlie our approach, as well as
observational and theoretical evidence that supports them. We then
proceed to give some general considerations in (\S
\ref{sec:gral_consid}) in both the non-magnetic and 
magnetic cases. With respect to the
former, in \S \ref{sec:non_mag_quali} we present a 
qualitative discussion of the reasons why hydrostatic configurations
are unlikely, in single-phase, nearly isothermal,
non-magnetic turbulent flows. This implies 
that dynamically compressed regions must either proceed to
collapse right away or re-expand. In \S \ref{sec:re-exp} we then
give a simple estimate of the re-expansion
time for clumps that do not collapse (which we call ``failed'' clumps).
In the magnetic case,  in \S \ref{sec:magn_quali} we give a
qualitative discussion of the formation and nature of cores in 
relation to the magnetic criticality of their parent cloud. 

In \S\S \ref{sec:numerics} and \ref{sec:num_results} we then present a
series of three-dimensional numerical 
simulations aimed at studying the lifetimes of cores in non-magnetic and
magnetic sub- and
supercritical environments, first describing the numerical method and
parameters (\S \ref{sec:num_meth}), resolution considerations
(\S \ref{sec:resol_consid}) and core analysis procedure (\S
\ref{sec:core_analysis}), and then the results (\S \ref{sec:num_results}). 
Finally, in \S \ref{sec:discussion} we discuss some implications and
caveats of our results, and in \S \ref{sec:conclusions} we give a
summary and concluding remarks.

\section{Assumptions} \label{sec:assumptions}

In this section, we describe the assumptions underlying the discussions
and calculations in this paper, as well as the evidence supporting them.
Nevertheless, as will be clear from the discussion, the
assumptions are not conclusively proven, and therefore it must
be kept in mind that our results are of course subject to the
applicability of our assumptions.

\subsection{Single-phase isothermal medium} \label{sec:strict_poly}

Throughout this paper we assume that the gas within molecular clouds
responds isothermally to compressions (i.e., that it obeys an isothermal
``effective'' equation of state $P=c^2 \rho$, where $P$ is the thermal
pressure, $\rho$ is the density, and $c$ is the sound speed), and that
it is at the same temperature throughout the volume considered. Although
this is the most common assumption for numerical simulations of
molecular clouds (see, e.g., the work of Padoan, E. Ostriker, Mac Low,
Klessen, Bate, Bonnell and their coworkers; see also the reviews by \VS\
et al.\ 2000; Mac Low \& Klessen 2004) because molecular gas is
typically at temperatures $\sim 10$--15 K, the true thermodynamic
behavior of molecular gas is still an open issue. 

The cores in
the survey  of Jijina, Myers \& Adams (1999) span almost an order of
magnitude in temperature, from $\sim$ 5 to 40 K, although indeed with a
strong concentration between 10--15 K, and presumably owing in most
cases to direct heating by nearby stellar objects, rather than to $PdV$
mechanical heating. In the absence of direct stellar/protostellar
heating, the gas tends to have more uniform temperatures. Indeed, the 
subsample of cores with no embedded clusters of Jijina et al.\ has a
narrower temperature distribution, while a study of the dark cloud TMC-1 by
Pratap et al.\ (1997) gives nearly uniform temperatures, between 8
and 10 K, even though the density spans nearly an order of magnitude.

Nevertheless, precise isothermality is not expected in molecular
clouds. Scalo et al.\ (1998) summarized results on the likely values of the
exponent $\gamef$ that appears in polytropic equations of
state of the form $P \propto \rho^{\gamef}$ (see also Myers 1978; Larson
1985). For molecular gas, Scalo et 
al.\ found $1/3 \lesssim \gamef \lesssim 1$ for $10^3~{\rm cm}^{-3}
\lesssim n \lesssim 10^5~{\rm cm}^{-3}$ in the absence of embedded
protostellar  sources. Recent calculations of the effective polytropic
index by Spaans \& Silk (2000) find a similar range for present-day
Galactic molecular gas.
More recently, Li, Klessen \& Mac Low (2003) have
investigated the dependence of the fragmentation process fragmentation
on the value of the polytropic exponent in molecular clouds. 

For the purposes of the present paper, the most important thermodynamic
consequences of assuming isothermality (i.e., $\gamef =1$) are that the
increase in thermal pressure during gravitational collapse
cannot eventually halt it (something which would require $\gamef > 4/3$
for spherically-symmetric collapse; e.g., Hayashi
1967; Larson 1969; Low \& Lynden-Bell 1976; Masunaga \& Inutsuka 1999;
see also the discussion in \S \ref{sec:non_mag_quali}),
and that the medium consists of a single thermodynamic phase. A
two-phase, or ``thermally bistable'', nature would require the existence
of an intermediate thermally unstable density range with $\gamef < 0$ between
density ranges with $\gamef > 0$ (e.g., Field, Goldsmith \&
Habing 1969). However, we see that the
$\gamef$-range estimated by Scalo et al.\ (1998) and Spaans \& Silk
(2000) for the densities of 
interest, $1/3 \lesssim \gamef \lesssim 1$, is completely contained
within the interval $0 < \gamef < 4/3$, and therefore 
also does not allow for neither of
those two phenomena. Thus, in this sense, our isothermal
assumption is consistent with those $\gamef$-range estimates. 

However, we must point out that Gilden (1984) has
suggested that a thermally unstable range with $\gamef <0$ may exist in
molecular gas, 
and, in consequence, allow for the possibility of thermal bi-stability.
If this does indeed occur, then our discussion of the
impossibility of producing Bonnor-Ebert-type hydrostatic structures in
\S \ref{sec:non_mag_quali}, based on the non-availability of a hotter,
more tenuous confining medium, is invalidated. Moreover, Blitz \& Williams
(1999) have argued that the HI gas may 
pervade the volume within clumps, by showing an anticorrelation between
the CO and the HI emission. This could also provide a confining medium
for BE-type structures. Ultimately, observations and
detailed studies of the heating/cooling processes in the gas should
determine the goodness of the isothermal assumtion for molecular
gas. 

Other relatively minor consequences of using an isothermal prescription
are that sound waves cannot steepen into shocks, and that the production
of vorticity through the baroclininc term is inhibited. The latter
limitation extends to polytropic prescriptions in general (see, e.g., \VS\
et al.\ 1996). This suggests that forthcoming improved molecular
cloud simulations should solve the internal energy equation, with 
reasonably realistic prescriptions for the cooling and heating processes
within them, as has been done for simulations of the global ISM (see,
e.g., the review by \VS\ 2002).

\subsection{Driven turbulence regime} \label{sec:driven_turb}

Another assumption underlying our work in this paper is that the
turbulence in molecular clouds is driven, rather than decaying, as we
have used almost exclusively simulations of driven turbulence.
Observationally, this is motivated by the fact that turbulence is
ubiquitous in molecular clouds (e.g., Larson 1981; Blitz 1993; Heyer \&
Brunt 2004),
including starless ones such as the so-called Maddalena's
cloud (Maddalena \& Thaddeus 1985). This suggests a universal origin and
maintenance mechanism for the clouds' turbulence (Heyer \& Brunt
2004). It has been suggested that
the turbulence in the clouds may arise from the very accumulation
mechanism that forms them out of the more diffuse atomic gas (\BP\ et
al.\ 1999a; Koyama \& Inutsuka 2002; 
\VS\ et al.\ 2003a,b; see also Hunter et al.\ 1986; Walder \& Folini
1998; \VS\ 2003), through dynamical
instabilities in the compressed regions.  In this case, the turbulence
in molecular clouds is driven for as long as the accumulation
process continues, while they probably disperse afterwards, as in the
recent simulations by Clark \& Bonnell (2004). 

In this scenario, the
clouds are in fact part of the turbulent cascade from the largest
scales down to the dissipative scales in the ISM, and the driving
therefore occurs from the large scales. This is supported by
recent comparisons between observations and either numerical simulations
(Ossenkopf \& Mac Low 2002) or fractional-Brownian-motion velocity
fields (Brunt 2003), in which the observed fields are most consistent
with the large-scale-driven turbulence numerical cases. Moreover, using
Principal Component Analysis, Brunt \& Heyer (2002) have estimated the
spectral index of the turbulence in 23 fields of the FCRAO CO survey of
the Outer Galaxy, finding that it is most consistent with the index 
found in numerical simulations of driven turbulence.
Thus, we assume that the appropriate model
for molecular clouds is that of driven turbulence, with injection at
the largest scales. 

In this paper we use the standard prescription of driving the flow
in Fourier space. Although this allows precise control of spectral
properties of the turbulence, such as the driving scale, it has the
downside of injecting the energy everywhere in physical space, and is
therefore not the most realistic energy injection mechanism. More
realistic simulations of molecular clouds should include physical
driving mechanisms, which, according to our discussion here, amount
possibly to the cloud formation process itself. In any case, the driving
scheme we use here should not directly
interfere with the cores' evolution, as it is applied on scales at least
10 times larger than typical core sizes. Specifically, the driving is
applied at scales of 1/2 the box size, while clumps and cores in the
simulations have typical sizes 0.08 and 0.04 times the box size (0.3 and
0.15 pc), respectively.

\subsection{Neglect of ambipolar diffusion} \label{sec:no_AD}

The numerical simulations we present in \S \ref{sec:num_results} are of
ideal MHD turbulence (subject only to the intrinsic numerical diffusion
of the numerical scheme), and do not include ambipolar diffusion (AD). This
implies that, in the case of subcritical simulations, gravitational
collapse is completely suppresed. In this regard, the recent simulations
by Li \& Nakamura (2004) including AD, give a more accurate description
of the effect of AD. On the other hand, those authors
restricted their simulaions to two dimensions and did not consider the
possibility of supercritical clouds, which we do here. 

As a zeroth-order approximation to determining the effect of AD in our
subcritical simulation, in \S \ref{sec:subcrit} we estimate the
classical AD time scale $\tad$ under the conditions of the clumps that form
there, which can then be compared to the clumps' lifetimes. If $\tad$ is
much larger than the latter, then AD is not expected to have
time to act during the cores' duration. If it is comparable or smaller,
then it can ``capture'' a transient, subcritical core, and increase its
mass-to-flux ratio, eventually causing its becoming supercritical and
collapse, although on time scales comparable to the dynamical lifetimes
of the clumps we find in the simulations, thus not significantly
delaying the collapse time.

\section{General Considerations}  \label{sec:gral_consid}

\subsection{The non-magnetic case}
\label{sec:non_mag_quali}

In this section we suggest that the
formation of hydrostatic configurations is unlikely in single-phase,
nearly isothermal (or, more generally, with $0< \gamef < 4/3$),
non-magnetic flows. This is because, for 
nearly-isothermal flows, the production of significant density fluctuations
requires supersonic turbulent compressions.
These fluctuations are in general transient, and so
the density enhancements they produce (``clumps'')
must also be so, 
unless the latter happen to ``land'' near a {\it stable} (attracting, 
in the language of dynamical systems) hydrostatic equilibrium between
self-gravity and thermal pressure (not turbulent pressure, because it is
locally too highly variable for establishing a static configuration). In
this case, after the external turbulent compression subsides, the core
may evolve 
towards this equilibrium rather than towards either re-expansion or
gravitational collapse.  

It follows that hydrostatic cores can occur only if there exist {\it
stable} equilibria. The best known such configurations in the
non-magnetic, isothermal case are the well-known Bonnor-Ebert
(BE) solutions of the Lane-Emden equation. However, the stability of
these configurations relies on the fact that they are truncated at a
finite radius from their center, being confined by a hotter,
negligible-density medium. The boundary between the cold, dense medium
and the hot, low-density one amounts to a transition between two stable
phases (with $\gamef >0$) mediated by a (non-populated) thermally
unstable density range with $\gamef < 0$.  

Of course, a hot confining medium is not necessary in general. For
polytropic situations with $\gamef > \gamma_{\rm c}$, where
$\gamma_{\rm c} = 4/3$ (Chandrasekhar 1961, \S 117; Yabushita 1968;
McKee et al. 1993, sec.\ VI.C; V\'azquez-Semadeni, Passot \& Pouquet
1996), spherically-symmetric equilibria can be found even for a
vanishing external pressure.\footnote{Note that for the realization of two- and
one-dimensional equilibria (i.e., with cylindrical and plane-parallel
geometries, respectively), the values of the critical polytropic
exponent are $\gamma_{\rm c} = 1$ and $\gamma_{\rm c} = 0$,
respectively (Ostriker 1964, 1965; Larson 1985; \VS\ et al.\
1996). However, for the three-dimensional, non-magnetic case 
we are considering here, we do not expect such configurations to be
applicable, as nothing locally restricts the flow to a lower
dimensionality. For example, filament production is often observed in
simulations, 
but collapse then proceeds {\it along} the filament (e.g., \BP\ 2004;
Klessen \& \BP\ in preparation), not perpendicularly
to it as would correspond to cylindrical geometry. On the other hand, in
the presence of a magnetic field, an anisotropy arises which tends to
make the turbulent compressions more one-dimensional. This is why stable
magnetostatic equilibria can exist in a sufficiently strongly magnetized
case, even though contraction can proceed freely along field lines.} 
This is why stars can be formed as stable entities from highly
anisotropic, dynamic, time-dependent accretion (Hartmann et al.\
2001). However, for bounded systems with $\gamef < 
4/3$, the boundary pressures are 
indispensible in establishing stable equilibria.\footnote{It is
interesting to note that Hunter (1977) found a class of {\it unstable}
BE spheres that do not collapse, but instead undergo large-amplitude
oscillations. However, these still rely on the presence of an external
confining medium at constant pressure and neglible density.} Such
boundaries are, however, ruled out in single-phase media,
in which no abrupt transition to a hotter, more tenuous medium at
the same pressure is possible.

We must examine, then, the possibility of {\it extended} (i.e.,
non-truncated) {\it stable} equilibrium configurations, in which the
core merges smoothly into its environment. To our knowledge,
this remains an open question. The best-known extended equilibrium
configuration is the singular isothermal sphere, which, however,
is known to be unstable, and has been used in fact as the initial
condition for gravitational collapse calculations (e.g., Shu 1977). A more
realistic (though still spherical) configuration would be one with a
{\it finite} central density. Such a configuration can be obtained by
numerically integrating the Lane-Emden equation to arbitrarily large
radii. However, this implies that it would be equivalent to a BE sphere
of arbitrarily large radius, and would thus be gravitationally
unstable\footnote{Again, in practice, the typical core-to-cloud mean
density ratio of $\sim$ 30--100 is
well into the unstable regime for BE-spheres.} (we thank D.\ Galli for
suggesting this argument). 

Other possibilities exist, however. Curry (2000) has found a new class
of periodic equilibrium structures in cylindrical geometry, which he
has proposed as candidates for the formation of cores in filamentary
clouds. However, he did not give definitive results concerning the
stability of these structures, and in fact concluded that full
numerical simulations may be necessary to resolve the issue.

In this respect, numerical simulations of non-magnetic
self-gravitating polytropic turbulence with $\gamef \leq 1$
suggest that hydrostatic configurations do not form: the turbulent
density fluctuations either re-expand, or proceed directly to
collapse. This was first shown by \VS\ et al.\ (1996) for
two-dimensional turbulent flows with $\gamef \leq 0.3$,\footnote{In this
2D case, $\gamma_{\rm cr} = 1$.} in which many generations of clumps
were seen to appear and disappear in roughly their crossing time,
until finally a strong enough compressive 
event occurred as to produce a gravitationally unstable core, which
then proceeded to collapse in roughly its own free-fall time. 
This result has subsequently continued to be found in 3D simulations of
non-magnetic, isothermal (e.g., Klessen et al.\ 2000) as
well as polytropic (Bate, Bonnell \& Bromm 2002; Li, Klessen \& Mac Low 2003)
turbulence, in which the formation of hydrostatic objects is in general not
observed (R.\ Klessen, private communication) for $\gamef < 4/3$. The turbulent
density fluctuations either re-expand or proceed directly to collapse
in their own free-fall time.

Finally, it is worth noting that Clarke \& Pringle (1997) have pointed out 
that cores cool mainly through optically thick lines, but are heated by
cosmic rays, and therefore may be dynamically unstable, as velocity
gradients may enhance local cooling. 


\subsection{Re-expansion time of ``failed'' compressions} \label{sec:re-exp}

The arguments above suggest that a significant fraction of the clumps
and cores in molecular clouds may be headed towards re-expansion rather
than collapse. This possibility is often
overlooked in the literature, which has generally focused on the
gravitational collapse of cores to form stars, and considers starless
cores as ``pre-stellar'' (with a few
notable exceptions; e.g., Taylor, Morata \& Williams 1996). However,
in the scenario advocated in this paper, a fraction of
the starless cores may never actually form stars, and instead
re-expand and disperse back into their parent molecular cloud, being
natural candidates to correspond to at least some of the observed
starless cores.

In the absence of self-gravity, one would expect that the re-expansion
of a spherical core should be nearly identical to the compression that
produced it, run backwards in time.
However, in the self-gravitating case, the re-expansion process must 
occur on a longer time scale than the compression because of the retarding
action of self-gravity. It is thus of interest to
estimate their expected lifetimes. 

A crude estimate of the re-expansion time can be given in terms of the
Virial Theorem (VT), similarly to the analysis performed by Hunter
(1977, sec. III.) to investigate the classes of solutions of unstable
BE spheres. We extend his treatment to estimate the 
re-expansion time scale, although we consider the case of zero
external pressure, because a) the density contrast between cores and
their parent molecular clouds are large enough (up to factors $\sim
100$) that the external thermal pressure is negligible during the
initial stages of the evolution with which we are concerned, and b)
in our case of an extended core, with a smooth density
profile, the whole profile varies in time, and therefore the pressure
is also variable at any boundary that we choose to define (which, as a matter
of fact, would be rather arbitrary). The full time-dependent problem is
what is solved by the numerical simulations, but for the purpose of a
simple estimate, here we consider the free
re-expansion of a gas sphere subject exclusively to its self-gravity
and internal pressure. 

\begin{figure}
\plottwo{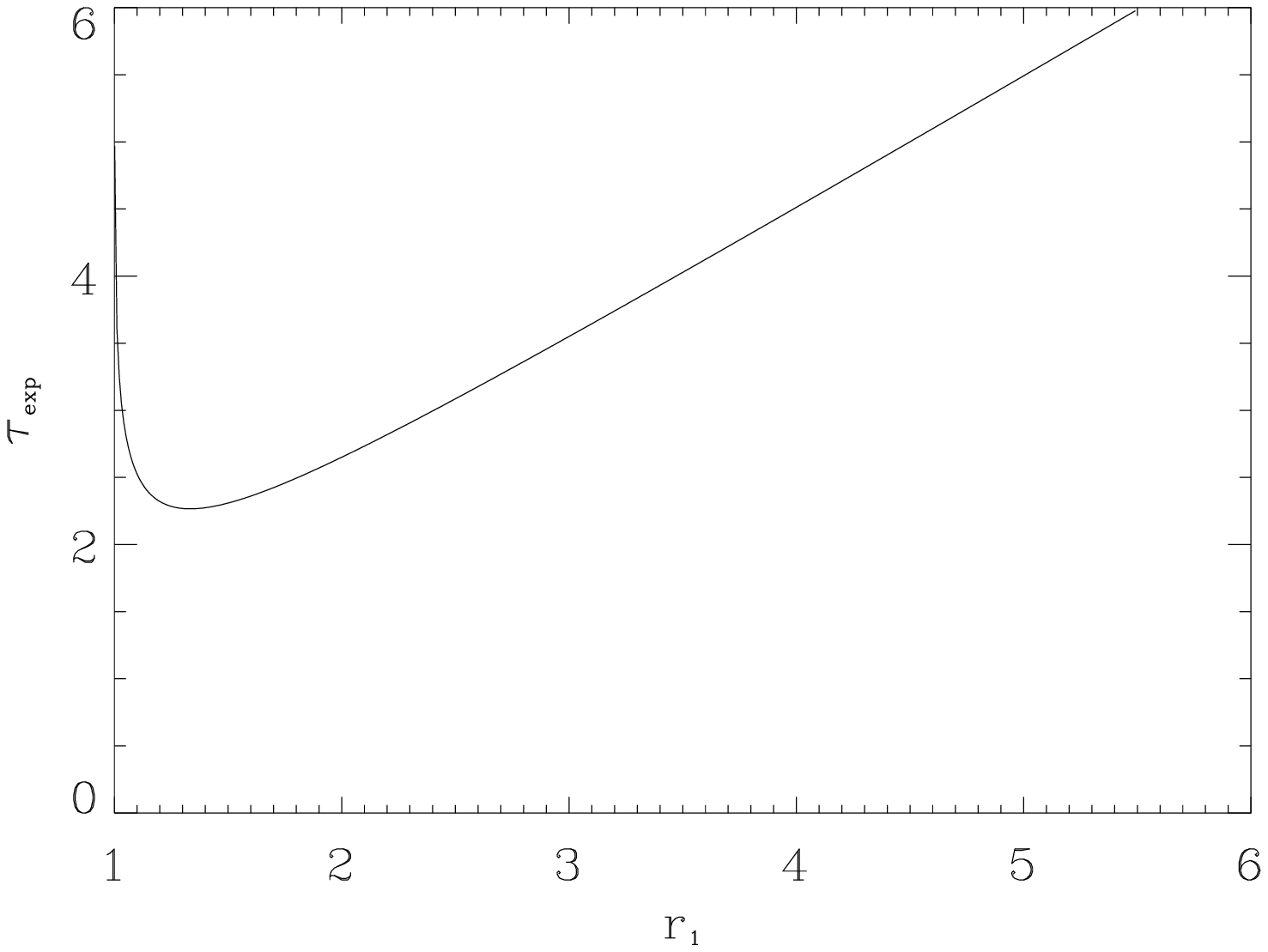}{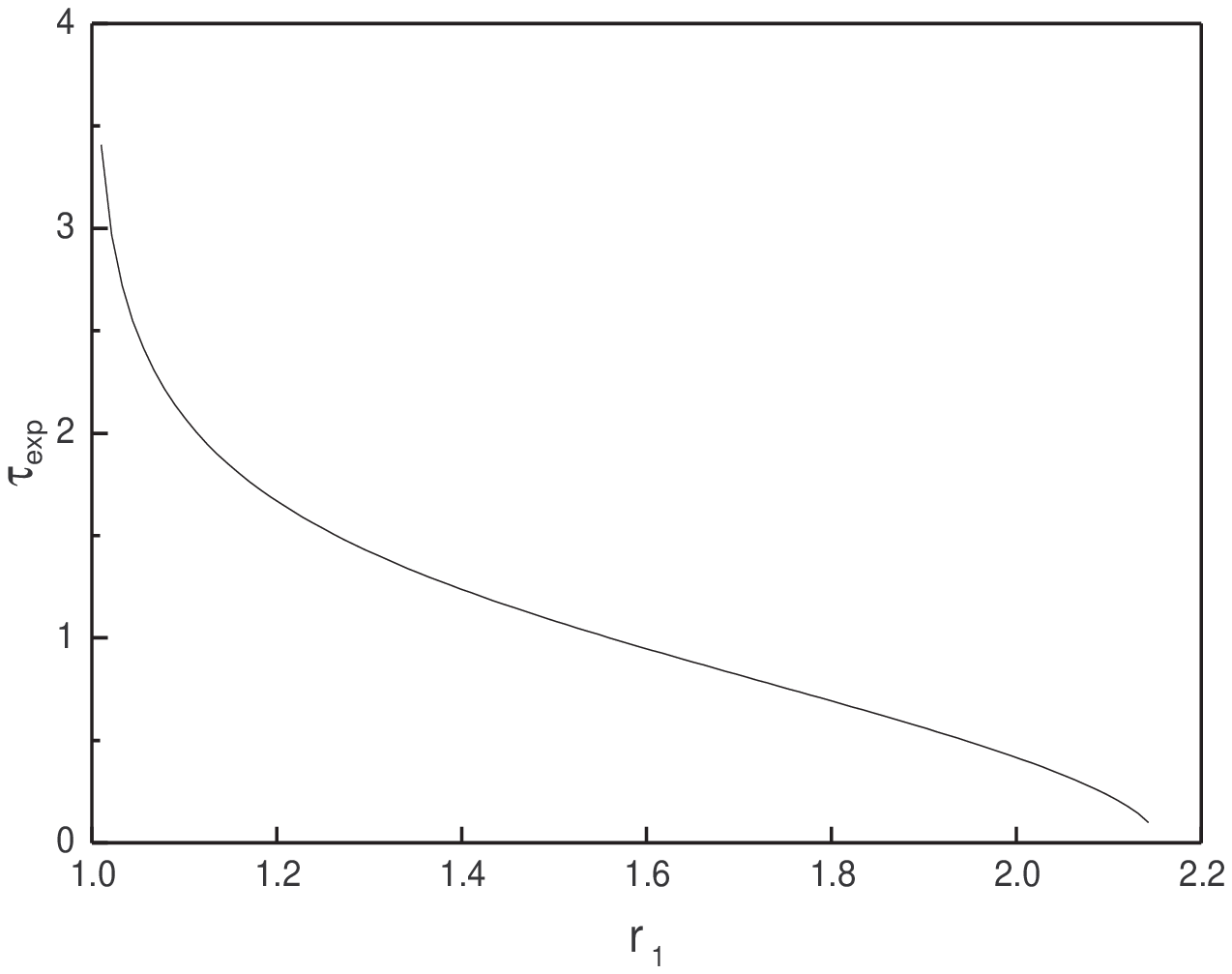}
\caption{ {\it a) (Left panel)} Re-expansion time of a core, in units of
the free-fall time, defined as the time necessary to 
double its initial radius, as a function of the initial radius
$r_1$, normalized to the equilibrium radius. {\it b) (Right panel)}
Simulated ``observer's perspective'' of the re-expansion time, giving
the time to go from the initial radius $r_1$ to a final one with density
$0.1~\rho_{\rm e}$, where $\rho_{\rm e}$ is the density of the equilibrium 
configuration. This emulates the time the core would be
visible to a high-density tracer sensitive to densities $>0.1\rho_{\rm e}$.} 
\label{fig:re-exp}
\end{figure}

The VT for an isothermal spherical gas mass
(``cloud'') of volume $V$ and mean density $\nm$ in empty space is
\begin{equation}
\frac{1}{2} \ddot I= 3 Mc^2 - \alpha GM^2/R,
\label{eq:VT}
\end{equation}
where the overhead dots indicate time derivatives, $M=\int_V \rho dV$ is
the cloud's mass, $R$ is its radius, $\rho$ is the density,
$I=\int_V \rho r^2 dV$ is its moment of inertia, $c$ is the
sound speed, and $\alpha$ is a factor of order unity. We can obtain an
evolution equation for the cloud's radius by 
replacing the radius-dependent density by its mean value in the
expression for $I$, to find $I\approx MR^2$. Thus, 
\begin{equation}
\frac{1}{2} \ddot I \approx M\left[R(t) \ddot R(t) + \dot
R^2(t)\right].
\label{eq:I_R}
\end{equation}
  From equations (\ref{eq:VT}) and (\ref{eq:I_R}), we obtain
\begin{equation}
\left[R(t) \ddot R(t) + \dot R^2(t)\right] = 3c^2 -\alpha GM/R.
\label{eq:R_evol}
\end{equation}
This equation can be integrated analytically, with solution
\begin{eqnarray}
\tau = & \frac{1} {\sqrt 3}\biggl[\sqrt{(r_2-1)^2-(r_1-1)^2} + \nonumber \\
 & \ln \left(\frac{r_2 -1+\sqrt{(r_2-1)^2 - (r_1-1)^2}}{r_1-1}\right) \biggr]
\label{eq:sol_R_evol}
\end{eqnarray}
where $r_1$ and $r_2$ are the initial and final radii of expansion,
normalized to the
equilibrium radius $R_{\rm e}=\alpha GM/3 c^2$, and 
$\tau=t/t_{\rm ff}$ is the time, non-dimensionalized to the
free-fall time $t_{\rm ff}=R_{\rm e}/c$. A characteristic
re-expansion time can be defined as the time required to double the
initial radius (i.e., $r_2 = 2 r_1$, a ``2-folding'' time), starting
from an initial condition 
$r_1>1$. Figure \ref{fig:re-exp}a shows this characteristic time as a
function of $r_1$. We see that when $r_1$ is very close to unity
(i.e., linear perturbations from the equilibrium radius), the
re-expansion time can be up to several times the free-fall
time. Values of $r_1$ moderately larger than unity have the shortest
re-expansion times, because the initial force imbalance is greater,
yet the final size is still not much larger than twice the equilibrium
radius. Finally, for larger initial radii (far from the equilibrium
value), the re-expansion time approaches that of free expansion at the
sound speed. These times are also long, but not because of a low
expansion velocity, but because the distance to be traveled by the clump
boundary from $r_1$ to $r_2$ increases proportionally to $r_1$. Figure
\ref{fig:re-exp}b shows an crude ``observer's perspective'' of the
re-expansion time, giving the time the core's density stays above 10\%
of the equilibrium-configuration density $\rho_{\rm e}$. This emulates
the time the core would be visible to a high-density tracer sensitive to
densities $>0.1\rho_{\rm e}$. This perspective clearly shows that the
longest re-expansion times of an object that can be called a ``core''
correspond to those that come closest to equilibrium.

We conclude that the re-expansion time is at least larger
than twice the free-fall time, making the probability of observing a
core in this process larger by this factor than that of observing a
free-falling core. This is consistent with the fact that molecular
clouds are generally observed to contain more 
starless than star-forming cores (e.g., Taylor et al.\
1996; Lee \& Myers 1999; see also Evans 1999 and references
therein).

A number of points are worth noting here. First,
the database put together by Jijina et al.\
(1999), reports a larger fraction of star-forming than starless cores
but, as those authors recognize in their section 7.1, their sample is
not free of biases, with starless cores not being as well represented
as the star-forming ones. This is due to the selection criteria used
to select the cores in the surveys from which these authors put their
database together. Instead, the Lee \& Myers (1999) dataset was
designed to be as free as possible from this bias, their cores being
selected optically from the STScI Digital Sky Survey. Second, 
some observations, such as those of ammonia cores, may be biased
towards selecting collapsing cores, if the densities they probe are very
large; i.e., if they probe densities not regularly reached by
non-collapsing clumps. The cores in the Jinina et al.\ sample, which is
an ammonia-core database, have larger typical number densities ($n >
10^4$ cm$^{-3}$) and a smaller starless fraction than those in the Lee
\& Myers optically-selected sample ($n \sim 7\times 10^3$
cm$^{-3}$). Indeed, in the 
simulations reported in \S \ref{sec:num_results}, the failed clumps
rarely exceed peak densities $\sim 5\times 10^4$ cm$^{-3}$.

\subsection{The magnetic case} \label{sec:magn_quali}

In the magnetic case, the classical Virial-Theorem (VT) analysis
(Chandrasekhar \& Fermi 1953; 
Spitzer 1968; Mouschovias 1976a,b; Mouschovias \& Spitzer 1976; Zweibel
1990) predicts the existence of sub- and super-critical 
configurations (e.g., Shu et al.\ 1987) depending on
whether the mass-to-magnetic flux ratio is below or above a critical
value $(M/\phi)_{\rm c}$. Subcritical configurations are known not to
be able to collapse gravitationally unless the magnetic flux is lost
by some process such as ambipolar diffusion. Supercritical
configurations, on the other hand, are analogous to the non-magnetic
case, except for the fact that the cloud behaves as if having an
``effective'' mass, reduced by an amount equal to the critical mass
(which depends on the magnetic field strength) (e.g., Spitzer 1968, \S
11.3.b). 

A magnetized clump or core (we generically use the term ``core'' in this
section) may thus be characterized by two nondimensional
numbers, the first one being its Jeans number ($J_{\rm c} \equiv
l/L_{\rm {J,c}}$), where $l$ is the characteristic core size and $L_{\rm
{J,c}}$ is the Jeans length at the mean core density, measuring whether
the core is gravitationally unstable with respect to the thermal
support ($J_{\rm c}>1$, or ``super-Jeans'') or
not. The other is the core's mass-to-magnetic-flux ratio expressed in units
of the critical value for collapse, $\mu_{\rm c} \equiv (M/\phi)/(M/\phi)_{\rm
cr}$. Collapsing cores must have $J_{\rm c}>1$ and $\mu_{\rm c} >1$,
while cores with $J_{\rm c}>1$ and $\mu_{\rm c} < 1$ can become
gravitationally bound 
but remain in a stable magnetostatic state (under ideal MHD
conditions). Finally, cores with $J_{\rm c}<1$  
are Jeans-stable, and must re-expand after a transient turbulent
compression, regardless of their value of $\mu_{\rm c}$. Note that both
$J_{\rm c}$ and $\mu_{\rm c}$ are in general time-dependent, and also
depend on the definition of the core's boundary. A fundamental
question is then what kind of cores are produced in turbulent clouds
as a function of the turbulent parameters and of the cloud's own sub- or
supercritical nature.


In order to obtain insight on this, it is instructive to consider the
formation of cores starting from idealized, initially uniform-density
clouds, truncated at a certain size, so that the cloud's mass-to-flux
ratio is well defined. This represents also the periodic-boundary
boxes of the numerical simulations. In reality, clouds are
embedded in the diffuse atomic medium, and in principle there is always
a large enough mass reservoir for them to be supercritical (e.g., Hartmann et
al.\ 2001), but if the region size that needs to be considered is very
large, then the time scale for accumulating this material in the
molecular region is long, during which the cloud can be considered
subcritical over the evolutionary time scales of individual clumps
within it. We thus consider both sub and supercritical uniform initial
conditions, under ideal MHD, and a fixed global mass-to-flux ratio.

In ideal MHD, it is clear that
an initially subcritical core can become supercritical only if its
parent cloud is supercritical. Indeed, the mass-to-flux ratio of an
{\it isolated} cloud or core is independent of its radius, because
both the mass and the flux 
are conserved quantities in this case. So, even if the whole cloud
were to be compressed by a turbulent fluctuation, it could not increase its
mass-to-flux ratio. As a consequence, a hard upper limit for the
mass-to-flux ratio of a core is that of its parent cloud. Hence the
need for the whole cloud to be supercritical if supercritical cores
can ever arise under ideal MHD. 

The above hard limit on the mass-to-flux ratio of a core can also be
seen as follows. Consider a cubic
cloud of size $L$ with uniform density $n_0$ and magnetic field
$B_0$, and any flux tube of square cross section $l^2$, with $l \ll
L$. Since in this case $M/\phi = N_0/B_0$, where $N_0 = n_0 L$ is the
column density across the cubic domain, it is clear that $M/\phi$ for
a flux tube is independent of its cross section, and equal to that of
the whole (uniform) cloud. If all the mass in
the flux tube were to end up in a single core, the core would still be
limited to having the same value of $M/\phi$ as its parent, uniform
cloud. If instead some fragmentation process (e.g., turbulent or
gravitational) causes the formation of several cores along the flux
tube, then each one of them will have a smaller value of $M/\phi$ than the
parent cloud, because the sum of their masses is bounded from above by
the mass of the flux tube, while the flux along the tube is constant. 

On the other hand, a lower estimate of the mass-to-flux ratio of a
core can be obtained by considering simply a region of volume $
l^3$ within the same uniform cloud. In this case, the mass-to-flux
ratio of this parcel (which is not properly a core, since it is at the
same uniform density as the whole cloud) scales as $l/L$. So, for a
core of size $l$ along the flux tube, we may expect its mass-to-flux
ratio $\mu_{\rm c}$ to lie in the range $(l/L)\mu_0 <\mu_{\rm c}<
\mu_0$ when its density $n_{\rm c}$ lies in the range $n_0 < n_{\rm c}
< (L/l)^3 n_0$, where $\mu_0$ and 
$n_0$ are respectively the parent cloud's mass-to-flux ratio and mean
density. 

%

\section{Numerical experiments} \label{sec:numerics}

\subsection{Method and parameters}  \label{sec:num_meth}

We use a total variation diminishing (TVD) scheme for solving the isothermal
MHD equations in three dimensions. TVD is a second-order-accurate
upwind scheme, and its 
implementation for isothermal flows is described in detail by Kim et
al.\ (1999). Here we use an extension of this code including
self-gravity and a random turbulence driver. The boundary conditions
in all our simulations are periodic. We solve the Poisson equation
using the usual Fourier method. In order to achieve second-order
accuracy in time, we implement an update step of the momentum density
due to the gravitational force as in Truelove et al.\ (1997). For the
turbulence random driver, we follow the method in Stone, Ostriker \&
Gammie (1998). As discussed in \S \ref{sec:driven_turb}, we drive 
the turbulence at large scales, at wavenumber
$k=2(2\pi/L)$, where $L$ is the one-dimensional size of the computational
box.  We adjust the kinetic energy input rate in order to maintain a
roughly constant specified rms sonic Mach number $\Ms$.


The simulations presented in this section can be considered 
extensions of previous  works by Pouquet, L\'eorat \& Passot (1991),
Ostriker et al.\ (1999), and Heitsch et al.\ (2001). In those studies, 
numerical simulations in which the entire computational domain is
supercritical systematically undergo collapse, albeit somewhat more
slowly than non-magnetic cases (by factors of 2--3). Moreover,  Li et
al.\ (2004) have shown that all cores arising 
in their supercritical simulations are supercritical by at least an
order of magnitude. Only when the
entire computational box is constrained to be subcritical is the
collapse of both the large and the small scales prevented, giving rise
to flattened structures. 
We extend those numerical works to study the lifetimes and criticality
of the individual cores in relation to the criticality of their parent
clouds. To this end, we now consider a series of 
simulations of Jeans-unstable numerical boxes,
considering both magnetically sub- and supercritical cases. In view of our
neglect of AD, AD-mediated contraction and subsequent gravitational
collapse cannot occur in our simulations. However, we can investigate
whether magnetostatic, subcritical cores with $J_{\rm c}>1$ and
$\mu_{\rm c} < 1$ (cf.\ \S \ref{sec:magn_quali}) form in the
simulations, that could then evolve 
quasi-statically under AD. This is done in \S \ref{sec:subcrit}.

The simulations are scale-free, and are characterized by three
non-dimensional numbers: $\Ms = \sigma/c$ (the rms sonic Mach number,
where $\sigma$ is the turbulent velocity dispersion and $c$ is the sound
speed), $J\equiv L/L_{\rm J}$ (the Jeans number, giving the 
size of the box in units of the Jeans length $L_{\rm J}$), and $\beta
\equiv P_{\rm th}/P_{\rm mag} = 8 \pi \rho_0 c^2/B_0^2$ (the plasma beta,
defined as the ratio of the thermal to the magnetic pressures, where
the subindex ``0'' denotes mean values in the box and $c$ is the
sound speed). The values of these parameters for all runs are
respectively given in
columns 2, 3 and 4 of Table \ref{tab:runs}, together with the run's
resolution (column 5) and
suitable physical units, obtained as follows: The sound speed is
assumed to be fixed, at $c=0.2$ km s$^{-1}$ (implying $T=11.4$ km
s$^{-1}$). Then, a 
Larson-like relation of the form $M_{\rm s} = 5 \left(L/1 {\rm
pc}\right)^{1/2}$ (e.g., Larson 1981; Blitz 1993) gives the box size
(column 6) as a function of $\Ms$ as 
\begin{equation}
L = 0.04\Ms^2~{\rm pc}.
\label{eq:L_Ms}
\end{equation}
The Jeans number and the box size give the mean number density density
(column 7) as 
\begin{equation}
n_0=500 \left[\frac{J}{\left(L/1~{\rm pc}\right)}\right]^2
{\rm cm}^{-3},
\label{eq:n_J_L}
\end{equation}
where we have assumed a mean molecular weight $m=2.4 m_H$.
In turn, the mean density and $\beta$ (at the above value of $c$)
give the mean field strength (column 8) as
\begin{equation}
B_0 = 0.205 \left(\frac{n_0}{\beta}\right)^{1/2} \mu {\rm G}
\label{eq:B_n_beta}.
\end{equation}
%
The last two columns in Table 
\ref{tab:runs} give the rms turbulent velocity dispersion and the
mass-to-flux ratio of the simulation (in 
units of the critical value), $\mu=(\beta/\beta_{\rm 
cr})^{1/2}$, where $\beta_{\rm cr}$ is given as a function
of $J$ as
\begin{equation}
\beta_{\rm cr} \sim 2/(\pi^2 J^2).
\label{eq:beta_crit}
\end{equation}
This in turn is obtained assuming that the critical mass-to-flux ratio is
$(M/\phi)_{\rm cr} \approx (4 \pi^2 G)^{-1/2}$ (Nakano \& Nakamura
1978). Specifically, we get $\beta_{\rm cr} 
\approx 0.023$ for $J=3$ and $\beta_{\rm cr} \approx 0.013$ for $J=4$. 
We have verified numerically the critical $\beta$ value for the $J=3$
case as follows. We set up a white-noise initial velocity field with
rms amplitude $10^{-4}c$ in each velocity component, and 
observe whether the maximum density in the simulation increases or
becomes saturated in time.  Indeed, the cases with $\beta \ge 0.03$
collapse in a runaway 
fashion, whereas the cases with $\beta \le0.01$ form magnetically-supported
configurations, flattened along the mean field direction.


\subsection{Resolution considerations} \label{sec:resol_consid} 

It is important to know when our results can be considered robust
physical ones rather than numerical artifacts caused by insufficient
resolution. A useful criterion for guaranteeing that the gravitational
collapse of an object is followed adequately was given by Truelove et
al.\ (1997). This criterion requires that the minimum 
Jeans length within the simulation be resolved by at least four zones.
Heitsch et al.\ (2001) increased this requirement to six zones for the
proper resolution of MHD waves, although this may be an excessive
constraint, as those authors showed that the magnetic field can only
prevent collapse if it is strong enough to provide {\it magnetostatic}
support. Our interest just being to determine whether
magnetostatically-supported structures arise, we consider that
fulfillment of the 4-zone Jeans condition is sufficient. This is
consistent with the results described below. 

In general, the constraint of resolving the Jeans length by
$N_{L_{\rm J}}$ zones limits the maximum resolvable density to
\begin{equation}
n_{\rm res} = \left(\frac{N_{\rm box}}{J N_{L_{\rm J}}}\right)^2 n_0,
\label{eq:Jeans_crit}
\end{equation}
where $N_{\rm box}$ is the number of zones per dimension in the
computational box. Note
that our definition of $J$ is different from that of Truelove et al.\
(1997). For
the Jeans criterion, $N_{L_{\rm J}} = 4$, so for our $J=4$ runs, this gives
$n_{\rm res} = 256 n_0$. For the more stringent criterion of MHD-wave resolution, $N_{L_{\rm J}} = 6$ zones, in which case $n_{\rm res} =
114 n_0$. 

\subsection{Core analysis procedure} \label{sec:core_analysis}

In the \S \ref{sec:core_props}, we describe 
in detail the evolution of the magnetic criticality and Jeans number
of some selected cores in the simulations. Here we describe the procedure for
measuring their mass-to-flux ratio $\mu_{\rm c}$ and Jeans number
$J_{\rm c}$. Note that this is not a fully 
unambiguous task, because the medium is a continuum. Thus, the cores'
boundaries, and therefore their masses and sizes, are somewhat
arbitrary. Moreover, the cores
are in general part of a larger parent clump, which is also involved in the
dynamics. A precise measurement of the structures' energy balance
would require to measure in principle all the terms in the virial
theorem for them, including the surface terms (\BP\ \& \VS\
1997; \BP\ et al.\ 1999a; Shadmehri, \VS\ \& \BP\ 2001; Tilley \& Pudritz
2004). Thus, here we
only measure the mass-to-flux ratio of the cores in an
approximate way, as follows. 

First, we define a structure (clump or core, which here we will
generically refer to as ``core'') as a connected region around a local
density peak, in which
the density is above a threshold value $\nt$. Then, the mass of the
core is computed directly from the density field as 
$M_{\rm c} = \int_{V_{\rm c}} \rho dV$, where $V_{\rm c}$
is the core volume. To compute the magnetic flux within the core, we
need in principle to first compute the (vector) average of the magnetic field 
($\langle {\bf B} \rangle$) within the core's volume, and then define a
surface $S$ normal to $\langle {\bf B} \rangle$ that bisects 
the core, in order to compute $\phi_{\rm c} = \int_S {\bf B} \cdot \hat
{\bf n} dA$, where $\hat {\bf n}$ is the unit vector normal to
$S$. For our purposes here, we use a simpler, approximate procedure. For
supercritical runs, in which the cores are not very far from round, we 
define the core radius as $R_{\rm c}=(3V_{\rm c}/4 \pi)^{1/3}$. It is
easy to show that, if instead of spherical, the core is a cylinder of
volume $\pi R^2 dz$, then an aspect ratio $a \equiv R/dz$ gives an error of a
factor of $a^{1/3}$ in the above estimate of the radius. At most, the
aspect ratios of the cores in our moderately supercritical run are 
$\lesssim 4$, and are even smaller in the strongly supercritical run. Thus,
the errors in the estimate of $R$ are by factors $\lesssim 1.6$. For the
subcritical run, and very flattened cores, we define the core radius as
the maximum distance between the density peak and all points belonging
to the core. Typical sizes for the ``clumps'' (defined by $\nt = 10
n_0$) and for the ``cores'' (defined by $\nt = 10 n_0$) are $\sim 0.08$
and $0.04$ times the box size ($\sim 0.32$ and 0.16 pc). 

With these definitions, we finally estimate the flux through the core is
$\phi_{\rm c} \approx \pi B R_{\rm c}^2$, where $B$ is 
the mean magnetic field intensity through the core. The core's Jeans number
is simply computed as $J_{\rm c}= R/L_{\rm J,c}$, where 
$L_{\rm J,c}=\left(\pi c^2/G \rho_{\rm c}\right)^{1/2}$, is the mean Jeans
length in the core, and $\rho_{\rm c}$ is the mean density in the core.


\section{Numerical Results} \label{sec:num_results}

\subsection{Global analysis}
\label{sec:global_analysis} 

\begin{figure}
\plotone{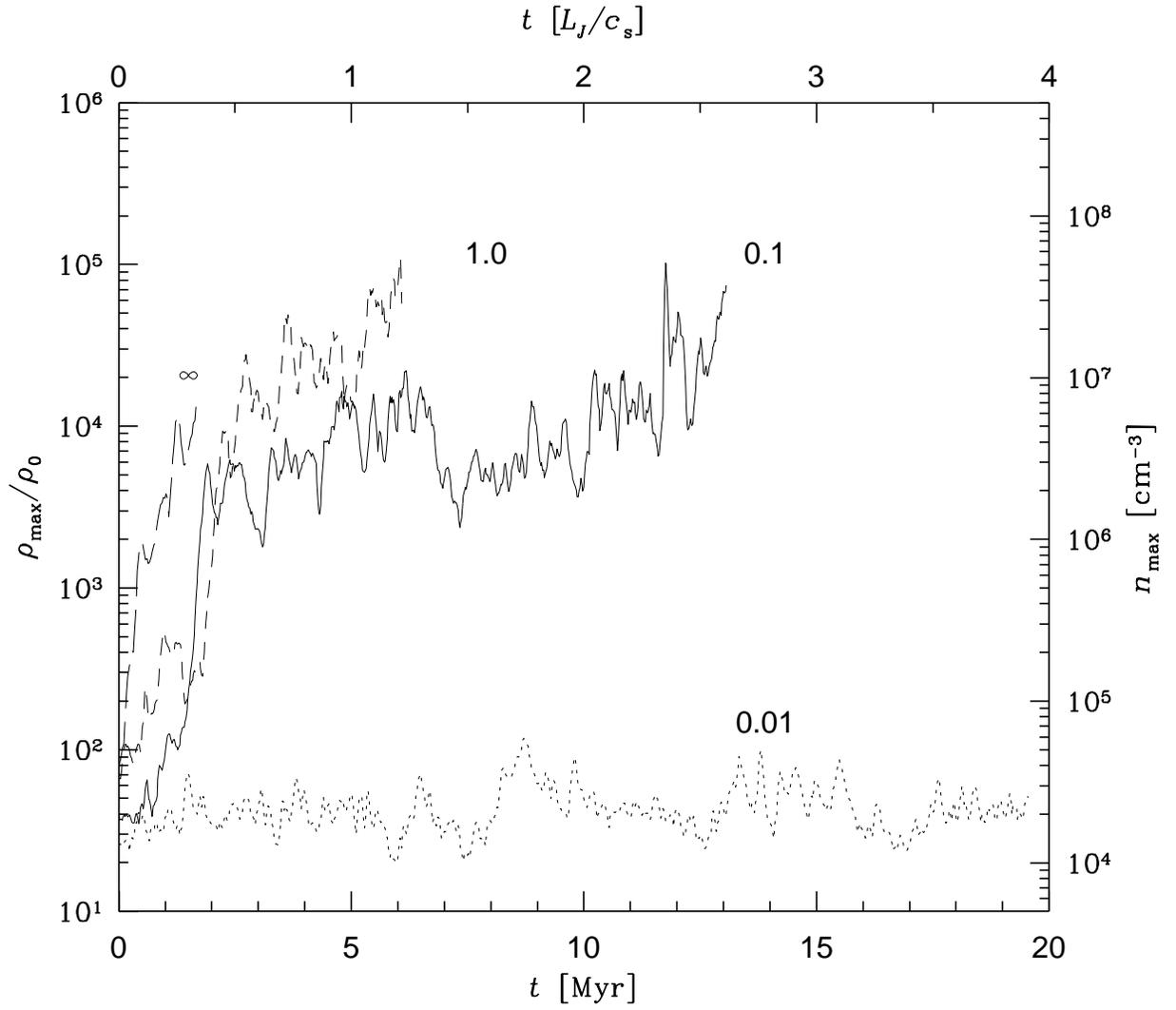}
\caption{Time-evolution of the global maximum of the density field for
runs M10J4$\beta$.01 ({\it dotted line}), M10J4$\beta$.1 ({\it solid
line}), M10J4$\beta$1 ({\it short-dashed line}) and
M10J4$\beta$$\infty$ ({\it long-dashed line}). The numbers indicate
the corresponding values of $\beta$. The upper time-axis is given in
units of the global free-fall time of the simulation.}
\label{fig:rho_max}
\end{figure}

Figure \ref{fig:rho_max} shows the evolution of the global density
maximum $n_{\rm max}$ for the four simulations in Table
\ref{tab:runs} with $\Ms=10$ and $J=4$ and $\beta = 0.01$, 0.1, 1 and
$\infty$, respectively labeled 
M10J4$\beta$.01, M10J4$\beta$.1, M10J4$\beta$1 and
M10J4$\beta$$\infty$. The time axis is shown both in Myr (lower axis)
and in units of the global free-fall time ($\tfg \equiv L_{\rm J}/c$,
upper axis). Note that runs M10J4$\beta$.01 and M10J4$\beta$.1 are
evolved for over 10 Myr. This is a
long time compared to recent estimates of cloud lifetimes of a few Myr
(\BP, Hartmann \& \VS\ 1999; Hartmann et al.\ 2001). The long integration
times are just for statistical purposes, and by no means imply that we
advocate longer lifetimes.

 From fig.\ \ref{fig:rho_max} it is
seen that in the subcritical ($\beta=0.01$) case,  $n_{\rm
max}$ is typically $\sim 40 n_0$ and seldom exceeds $\sim 100
n_0$. The same behavior is observed in the other subcritical runs (not
shown), M10J3$\beta$.01, M8J4$\beta$.01, and M10J4$\beta$.01LR.
Thus, in these simulations core-like densities ($n \gtrsim 200 n_0\sim
10^5$ cm$^{-3}$) are not produced. The fact that the maximum density
does not stay constant also suggests that there are no magnetostatic
clumps. Note that these results cannot be attributed to numerical diffusion,
whose effect would be expected to act in the opposite direction of
causing artificial flux loss and possibly collapse.

 In contrast, fig.\ \ref{fig:rho_max} also shows that in the
supercritical runs $n_{\rm max}$ reaches values $\sim 10^4 n_0$ after
roughly $1/2~\tfg$, although the rise from densities $\gtrsim 100 n_0$
to $\sim 10^4 n_0$ occurs on a much shorter time scale. This corresponds
to the first local collapse event, as discussed in \S
\ref{sec:core_analysis}. Finally, the non-magnetic run
M10J4$\beta$$\infty$ is seen to start producing collapsed objects much
earlier than the magnetic supercritical runs. Thus, {\it both the
supercritical runs and the non-magnetic ones produce the first collapsed
objects on times shorter than the global free-fall time} as a consequence
of the production of locally gravitationally unstable objects by the
turbulence. However, 
it is seen that the effect of the turbulent compressions is diminished
by the presence of the magnetic field even in the strongly supercritical
case in comparison to the non-magnetic one.

\subsection{Evolution of core properties} \label{sec:core_props}

\subsubsection{Subcritical case} \label{sec:subcrit}

\begin{figure}
\plotone{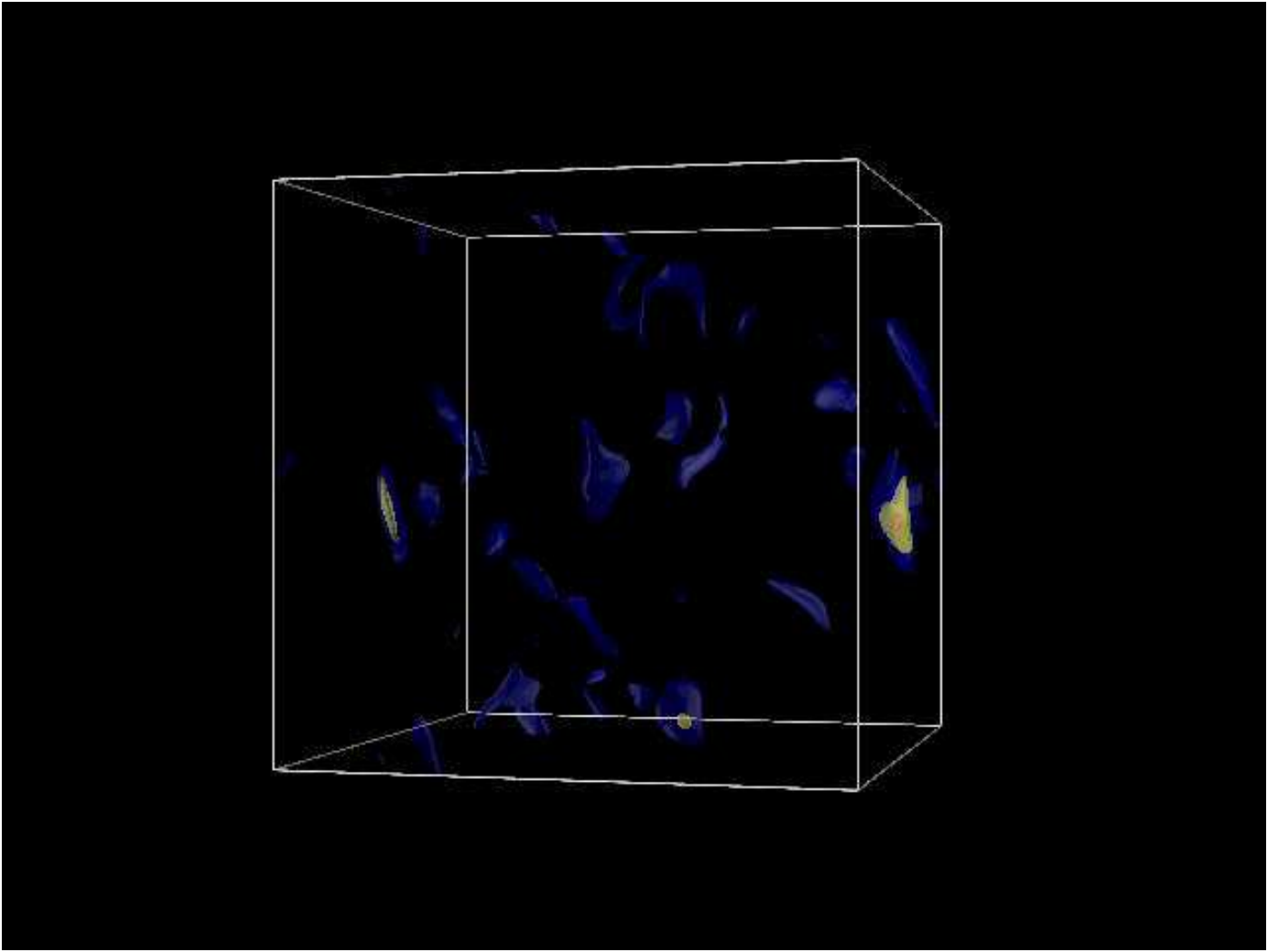}
\caption{Animation of the density field for the subcritical run
M10J4$\beta$.01, showing transparent density iso-surfaces at
$n = 10 n_0$ (blue), $n = 30 n_0$ (yellow),  and $n = 100
n_0$ (red). The figure shown in the paper edition corresponds to
frame 223, at which the moderately-long-lived core attains its 
time-wise maximum central density, of just over $100 n_0$.} 
\label{anim:sub}
\end{figure}

In fig.\ \ref{anim:sub}, we show an animation of the evolution of the
subcritical run 
M10J4$\beta$.01. Depicted are three isosurfaces at densities 10 (blue), 30
(yellow) and 100 (red) times $n_0$. The time interval
between successive frames is $\Delta t = 0.002~\ts$, where
$\ts = 20$ Myr is the sound crossing time for the $\Ms=10$ ($L=4$ pc)
boxes. In this 
simulation, most density peaks (``clumps'', in this case) are seen to
be extremely transient in nature, with lifetimes 0.01--0.075 $t_{\rm
s}$, or 0.2--1.5 Myr (some 5--30 frames), and with poorly defined
identities, the clumps actually merging with other structures or
simply morphing into different shapes and sizes. 

In this simulation, we only observe one clump that becomes
gravitationally bound 
and survives for a few local free-fall times. It appears near the left
boundary of the box around frame 205, and moves across it to remain
near the right boundary for the rest of its existence. This core
maintains peak densities $\np > 100 n_0 = 5\times 10^4$ cm$^{-3}$ during
7 frames, from frame 221 to 227 (0.28 Myr), and $\np \gtrsim 60
n_0$ during $\sim 28$ frames ($\sim 1.1$ Myr), from frame 210 to frame
237. Figure \ref{fig:mu_J_sub} shows the evolution of $\mu_{\rm c}$ and
$J_{\rm c}$ for this clump (defined by $\nt =40 n_0$) from frame 210 to
240 ($t=8.4$ to 9.6 
Myr). We see that it remains strongly subcritical (well below the value
for the box of 0.9), but it manages to remain significantly super-Jeans
for slightly over 1.2 Myr. 

This clump then represents the closest our subcritical simulation gets
to producing the conditions for subsequent quasi-magnetostatic
evolution mediated by AD. Nevertheless, we see that this clump's
lifetime is still quite short, because it is not very strongly
gravitationally bound, and the turbulence manages to tear it appart
after having formed it. It is then natural to ask whether AD would
have enough time, had it been included in the simulations, to allow it
to become more strongly bound, and ultimately collapse, before it is
dispersed again by the turbulent motions. 

To answer this question, Appendix I presents a simple estimate of
the classical AD time scale under the conditions of this clump. It is
found that $\tad \gtrsim 1.3$ Myr, which is very close to the duration of
this clump in our simulation, suggesting that AD is possibly capable
of ``capturing'' it and helping it decouple from the external
turbulence. Note, however, that this value of $\tad$ is not
significantly larger than the collapse times we observe in the
supercritical simulations, and thus a core that evolves on this time
scale can hardly be considered quasi-static. 

Another relevant question is whether such clumps would approach a
magnetostatic state if the turbulence were decaying. To test this
possibility, we have conducted the additional experiment of turning off
the turbulent driving at frame 210, right after the clump forms. In this
case, we find that indeed the clump undergoes a few bounces,
but maintains its identity for at least 
1.6 Myr (the duration of the integration we performed), with no sign of
disruption towards the end. This suggest that, if turbulence is turned
off right after it forms the clump, then the clump does evolve towards a
magnetostatic state, and can easily be captured by AD, to evolve in
accordance to the standard model. Presumably, other clumps might undergo
the same evolution in a decaying situation, as in the simulations by Li
\& Nakamura 2004.

We conclude from this section that even the densest,
longest-lived clumps arising in our subcritical simulation are
dispersed here because the strong magnetic field does not allow them to
collapse nor achieve very strong gravitational binding, and makes them
susceptible of subsequent disruption by the same turbulent
field. The longest clump lifetimes are marginally long enough for them to
undergo AD-mediated evolution, albeit on relatively short time
scales. However, such clumps could 
easily become magnetostatic in a decaying turbulent field. 
The fundamental question then becomes whether molecular clouds are
continually driven or decaying (see discussion in \S \ref{sec:discussion}).

\subsubsection{Moderately supercritical case}
\label{sec:mod_super}

\begin{figure}
\plottwo{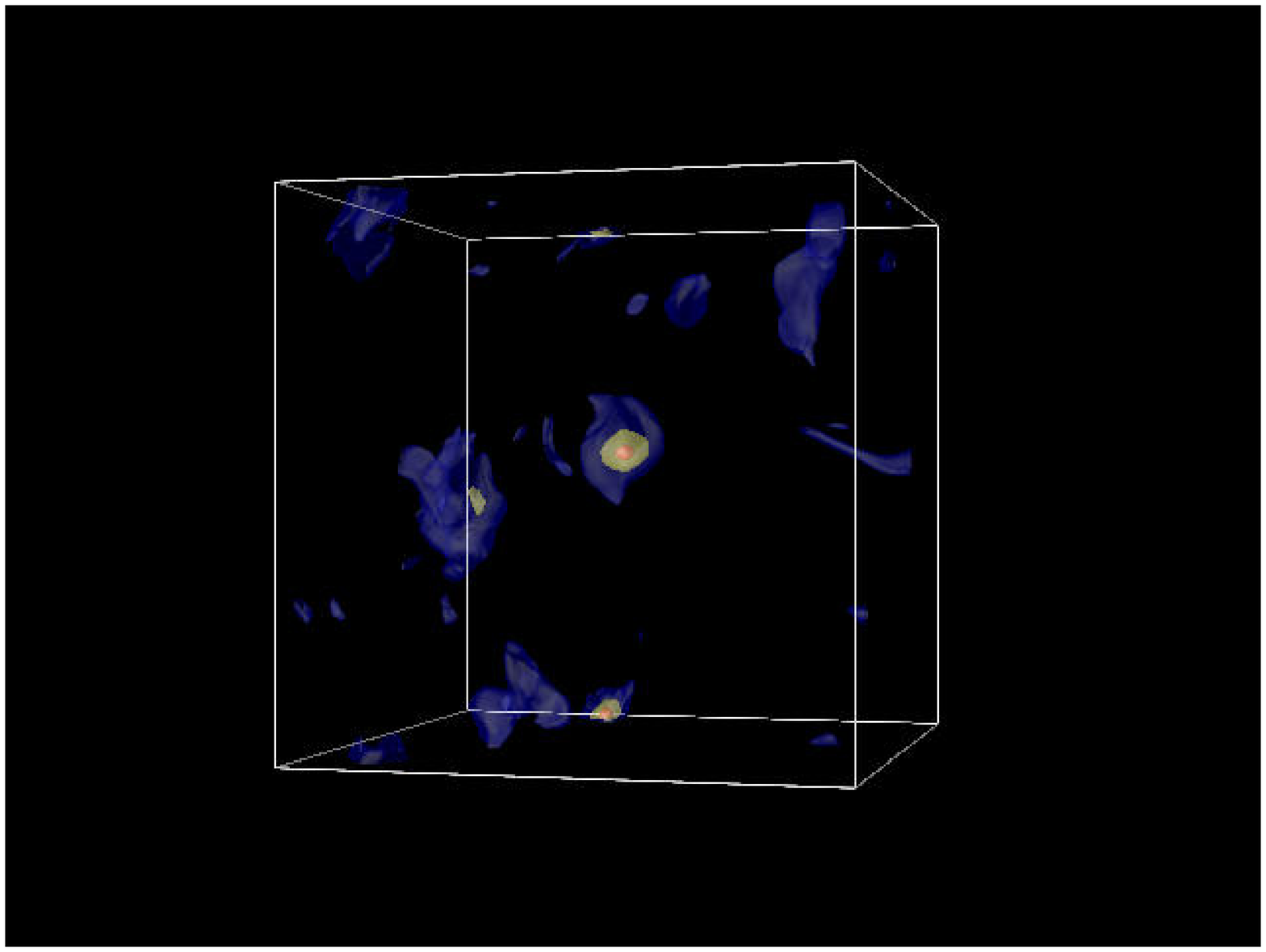}{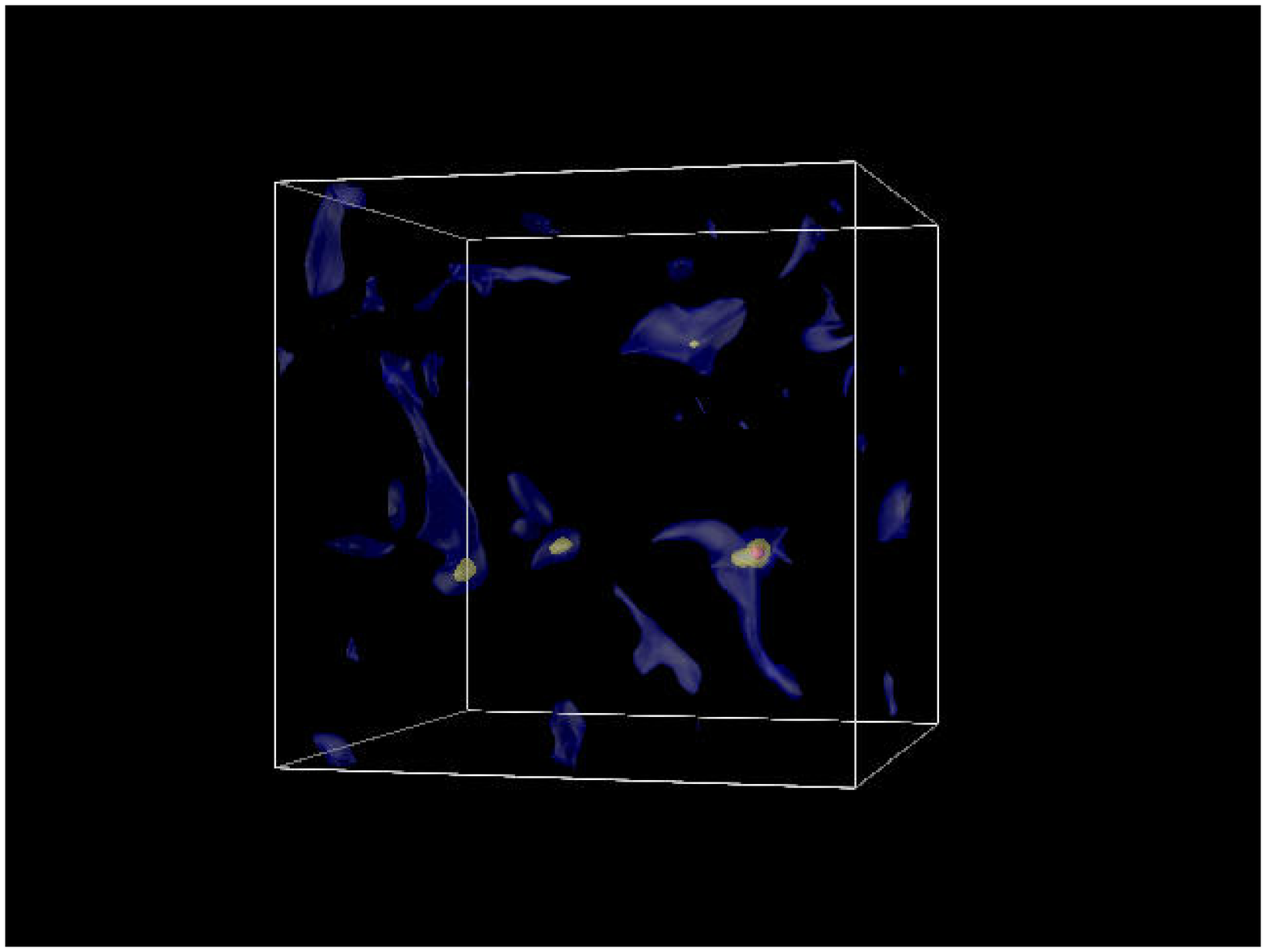}
\caption{Animations of the supercritical runs as in fig.\
\ref{anim:sub}, but with the density iso-surfaces corresponding to
$n = 10 n_0$ (blue), $n = 100 n_0$ (yellow),  and $n =
1000 n_0$ (red). {\it a)} Run M10J4$\beta$.1. {\it b)} Run
M10J4$\beta$1. The still frames in the paper edition show: {\it (a)}
Frame 210 of run M10J4$\beta$.1, at which time the third (and last)
collapsing core (on the left) is
forming. This core is fully collapsed by frame 220. {\it (b)} Frame 55 of run
M10J4$\beta$1, showing two of the long-lived but non-collapsing cores at
the center (detailed in fig.\ \ref{fig:mu_J_str_sup}) and left, and a
collapsed core on the right.}  
\label{anim:super}
\end{figure}

Figure \ref{anim:super}a shows the evolution of the moderately
supercritical run M10J4$\beta$.1, again with $\Delta t = 0.002~\ts$,
but now with the 
isodensity contours denoting 10 (blue), 100 (yellow) and 1000 (red)
$\times n_0$. It is seen that
in this run, objects reaching much higher {\it peak}
densities, $\np \gtrsim 
5000 n_0$, form since early in the evolution. Specifically,
three such objects form at times $\sim
1.76$, 5.2 and 8.4 Myr (frames 44, 130 and 210, respectively). Once
they have reached those densities, these objects are already collapsed,
as they simultaneously have $J_{\rm c} \gg 1$ and $\mu_{\rm c} \gg 1$,
and so nothing supports them against their self-gravity. In particular,
they have $\mu_{\rm c} > \mu_{\rm box}$, indicating that numerical
diffussion has caused magnetic flux loss within them. They do
not proceed to a singularity because, once most of the collapsing mass
is in a single, completely unresolved numerical cell,  the
collapse cannot be followed any further by the numerical grid. These
cells, which we refer to as ``sink cells'', can be thought of as the fixed-grid
equivalent of sink particles in SPH simulations (Bate, Bonnell \& Price
1995), and are standard fare in fixed-grid numerical studies (e.g.,
Heitsch et al.\ 2001; Li et al.\ 2004). Note that, however, the sink
cells continue to accrete from their surroundings at a slower
rate, and eventually the simulation finally stops
when the density gradient becomes too large. Nevertheless, as we show below,
the {\it onset} of collapse occurs under well-resolved conditions.

\begin{figure}
\plottwo{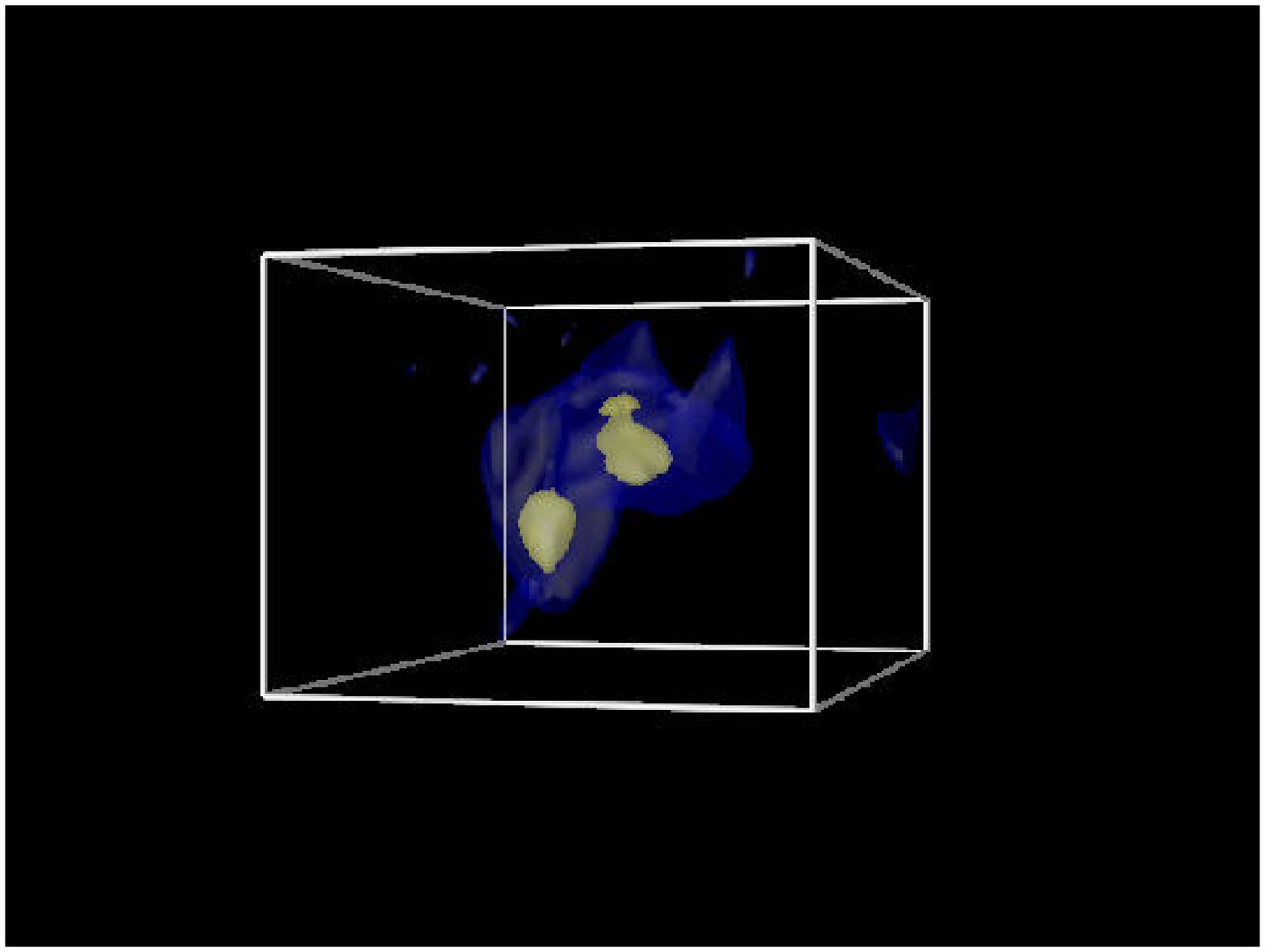}{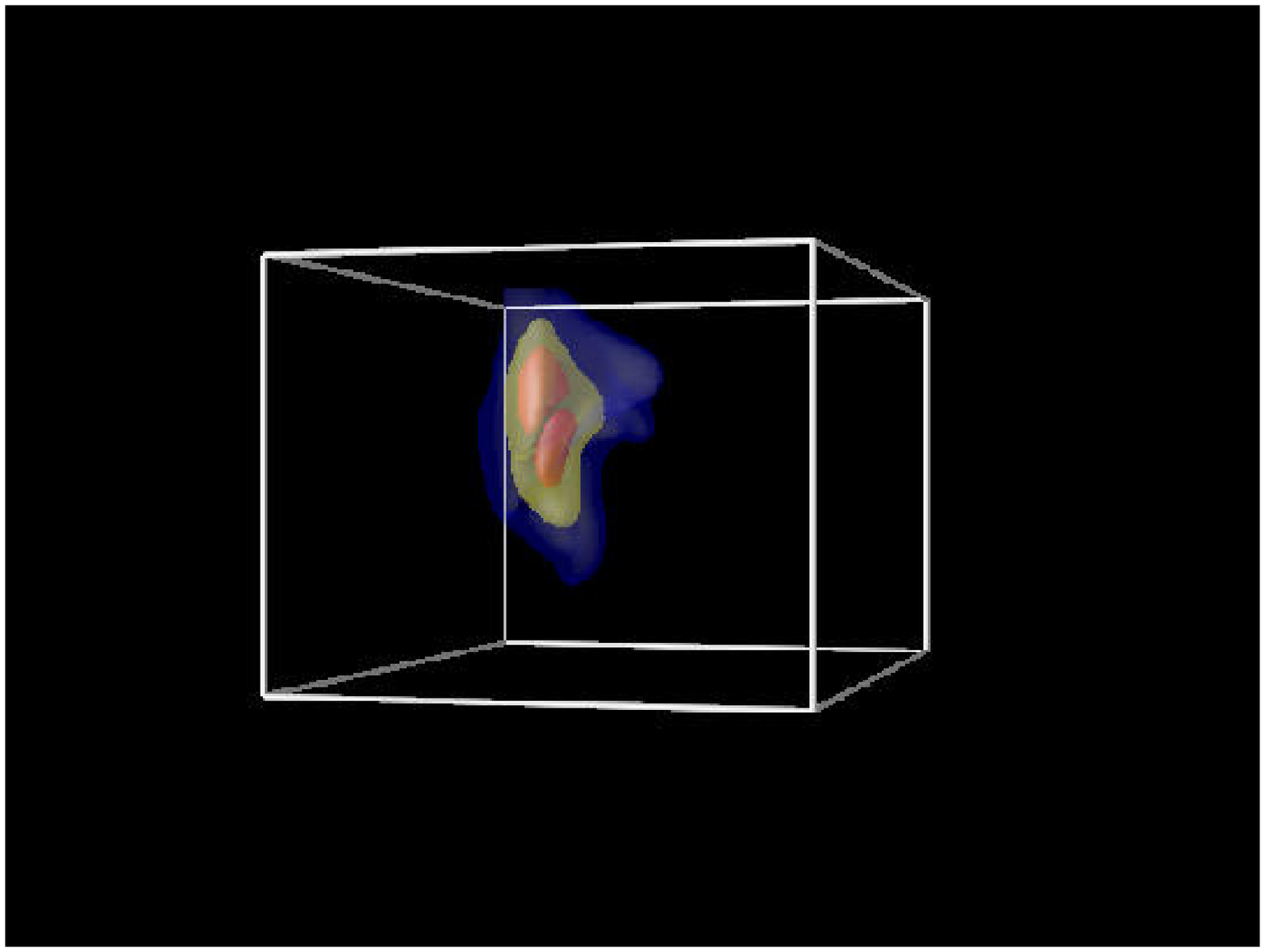}
\caption{Iso-density surface 3D map of the clump-core
system appearing at early times in run M10J4$\beta$.1, detailed in
fig.\  \ref{fig:mu_J_mod_sup}. The times shown are {\it a)} $t= 0.06 \ts$
(frame 30 in the animation of fig.\ \ref{anim:super}a ) and {\it b)} $t=
0.08 \ts$ (frame 40). The iso-density surfaces are at $n= 10 n_0$
(blue), $n= 40 n_0$ (yellow) and $n = 100 n_0$ (red). The latter does
not exist at frame 30.} 
\label{fig:clump_core_b.1}
\end{figure}

Now we focus on the evolution of the first collapsing object. From the
animation in 
fig.\ \ref{anim:super}a, we see that the collapsing core actually
forms from the merger of two previous cores, which had formed almost 
simultaneously at frame 36 within a clump located near the center of the
computational box. The two cores merge at frame 44, which marks the
``birth'' of the collapsing core, all happening within the larger clump. 
Thus, we investigate the criticality and stability of both the ``parent''
clump and the ``daughter'' cores at frames 30, 40 and 50.
Figures \ref{fig:clump_core_b.1}a and \ref{fig:clump_core_b.1}b respectively
show 3D iso-density surface maps of the clump/core  system in question
at frames 30 ($t=1.2$ Myr) and 40 ($t=1.6$ Myr). The iso-surfaces are
drawn at densities $n=10 n_0$ (blue), $n=40 n_0$ (yellow) and $n=100
n_0$ (red). 

At frame 30 (fig.\ \ref{fig:clump_core_b.1}a), it is seen that the
$n=100 n_0$ iso-surface does not exist, since the peak density is $<100
n_0$. It is also seen that the $n=10 n_0$ 
isosurface defines a single structure, which we refer to as the parent
``clump'', while the $n=40 n_0$ isosurface resolves the two cores. At
frame 40 (fig.\ \ref{fig:clump_core_b.1}b), the $n=40 n_0$ isosurface
does not resolve the two daughter cores anymore, although the $n=100
n_0$ isosurface still does. It is important to remark that the
isosurfaces drawn at a certain value of $\nt$ at different times {\it do
not} trace the same gas parcels; i.e., the isodensity surfaces are not
Lagrangian boundaries. Neither are clump boundaries defined by means of
standard clump-finding algorithms.

\begin{figure}
\plotone{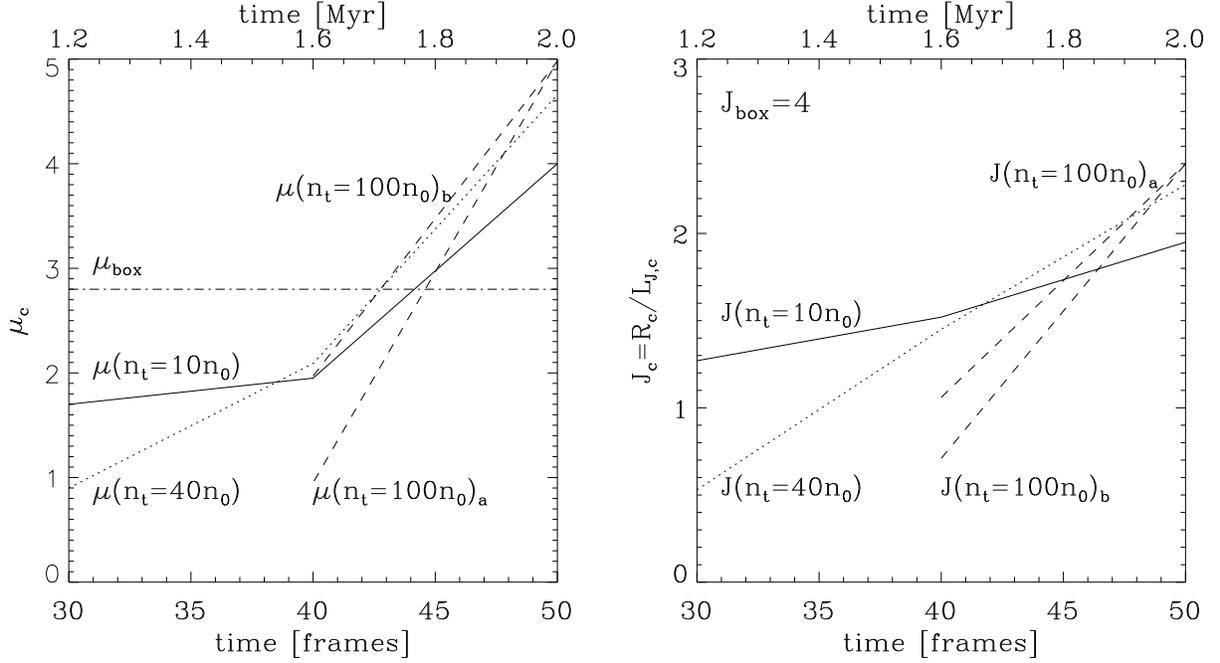}
\caption{Evolution of: ({\it Left panel}) the mass-to-flux ratio (in
units of the critical value) $\mu_{\rm c}$ and of ({\it right panel})
the Jeans number $J_{\rm c}$ for the first collapsing object in the
moderately supercritical run M10J4$\beta$.1. The various curves
correspond to different values of the clump-defining threshold density
$\nt$. Note that at frame 40 there are two cores defined by $\nt = 100
n_0$, described by the two dashed lines. At frame 30 the peak density is
below $100 n_0$ and therefore the dashed lines do not extend down to
this frame. Note also that at frame 30 there are two cores defined by
$\nt=40 n_0$, but their $\mu_{\rm c}$ and $J_{\rm c}$ are nearly
idenstical, and so only one curve is shown.}
\label{fig:mu_J_mod_sup}
\end{figure}

Figure \ref{fig:mu_J_mod_sup} shows the evolution of the mass-to-flux
ratio (panel {\it a}) and of the Jeans number (panel {\it b}) for this
collapsing 
object from frame 30 to frame 50 ($t=1.2$ to 2 Myr). Three curves are
shown in each panel, each corresponding to different values of the
clump-defining threshold density $\nt$. Note that two cores exist at
frame 30 for $\nt=40 n_0$, but their $\mu_{\rm c}$ and $J_{\rm c}$ are
nearly identical, and so they are described by the same (dotted)
line. Similarly, at frame 40 there are two cores defined by the $\nt=100
n_0$ level, and in this case their $\mu_{\rm c}$ and $J_{\rm c}$ are
different, so they are described by two separate (dashed) lines, which
in fact start at frame 40, because the cores did not exist at
threshold $\nt = 100 n_0$ at frame 30.

  From fig.\ \ref{fig:mu_J_mod_sup} it is seen that
that {\it the parent clump is already
Jeans-unstable and magnetically supercritical at frame 30}, even
though it contains two well-defined substructures, both of which
individually are 
subcritical and sub-Jeans at this time, and are therefore probably not
collapsing themselves at this time. In this sense, the merger of
the two cores can be considered a part of the global collapse of the
larger structure, a fact which is confirmed by noting that in the animation,
the size of the $n=10 n_0$ isosurface is also shrinking, implying that
most of the mass in the whole clump is involved in the collapse (see
also fig.\ \ref{fig:clump_core_b.1}).

Note that the super-Jeans and supercritical character of the $\nt=10
n_0$ clump at frame 30 cannot be attributed to numerical diffusion, as
its mean density is only $\sim 20 n_0$, with its peak only reaching
$83 n_0$, thus very safely below even the most stringent (and probably
excessive) resolution
constraint of Heitsch et al.\ (2001) of $\nres=114 n_0$. This is also
supported by the fact that at this time the mass-to-flux ratio of the
clump, albeit supercritical, is still smaller than that of the whole
box, in agreement with the constraint that no structure within the simulation
can have a value of $\mu$ larger than that of the whole box under ideal
MHD. The latter property holds also for the two cores defined by $\nt =
100 n_0$ at frame 40, whose 
mean and peak densities are respectively $\langle n_1 \rangle = 153
n_0$, ${\np}_1 = 337 n_0$ and $\langle n_2 \rangle = 121 n_0$, ${\np}_2 = 166
n_0$. It is only 
at frame 50, when the core has a peak density of over $5000
n_0$ and an average density of $329 n_0$ (above $\nt=40 n_0$), that its
mass-to-flux ratio exceeds that of the box, indicating that at this time
numerical diffusion has clearly caused significant loss of magnetic
flux. However, 
the collapse was initiated when the whole structure was cleanly
resolved, and therefore must be considered a robust physical
result. That is, {\it the formation of a supercritical, super-Jeans
structure causes collapse and ultimately numerical diffusion, rather
than the collapse being an artifact of numerical diffusion.} The same
happens during the formation and collapse of the other two collapsed
cores (at frames 130 and 210). This conclusion is
also supported by the fact that no collapse occurs in the subcritical
simulation, even though it is only marginally so ($\mu = 0.9$), which
shows that the simulations accurately do not allow collapse when it
physically should not occur. 


Finally, two more points are worthy of notice. One is that this first
collapsing object
goes from first appearance to a fully collapsed state in less than $2
\tfc$, where $\tfc \equiv L_{\rm J,c}/c$ is the local free-fall time in
the core, with $ L_{\rm 
J,c}$ being the Jeans length at the core's mean density. At a mean density
$\langle n \rangle_{\rm c} = 100 n_0$, $\tfc = 0.5 {\rm Myr} = 12.5$
frames. The collapse corresponds to the rapid rise of $n_{\rm max}$ 
from $\sim 10^2$ to $\sim 10^4 n_0$ in fig.\ \ref{fig:rho_max}. The
second is that, as in the non-magnetic case, 
moderate-density ($\np \gtrsim 10^2$ cm$^{-3}$) clumps arise that
redisperse on time scales 1/2--1 $\tfc$ ($\lesssim 10$ frames).

\subsubsection{Strongly supercritical case} \label{sec:strong_super}

In the strongly supercritical run M10J4$\beta$1 (fig.\
\ref{anim:super}b), collapsing objects undergo a very similar
evolution as in the moderately supercritical one, M10J4$\beta$.1.
However, this run exhibits an additional interesting phenomenon: Besides the
rapidly-collapsing dense cores and the rapidly-dispersing intermediate
density clumps discussed above, we also observe the formation of two
long-lasting, intermediate-density clumps ($\np$ 200--300 cm$^{-3}$)
which {\it do not collapse}, but which
take times $\sim 0.15~\ts$ ($\sim 0.6~\tfg$, $\sim 75$
frames, or $\sim 3$ Myr) to redisperse. These are seen as the
long-lasting clumps that do not have a red isodensity surface in the
animation. One of them is
formed at frame 16 ($t=0.64$ Myr) near the top of the simulation, and disperses
around frame 84 ($t=3.36$ Myr). The other one forms at frame 42
($t=1.68$ Myr) towards the left of the simulation, at roughly half the
box height, and disperses around frame 120 ($t=4.80$ Myr). These clumps
are marginally 
resolved according to the Jeans criterion, but nevertheless end up
{\it redispersing}, rather than collapsing, suggesting that numerical
diffusion of the magnetic field is not a concern yet in these
clumps.

\begin{figure}
\plotone{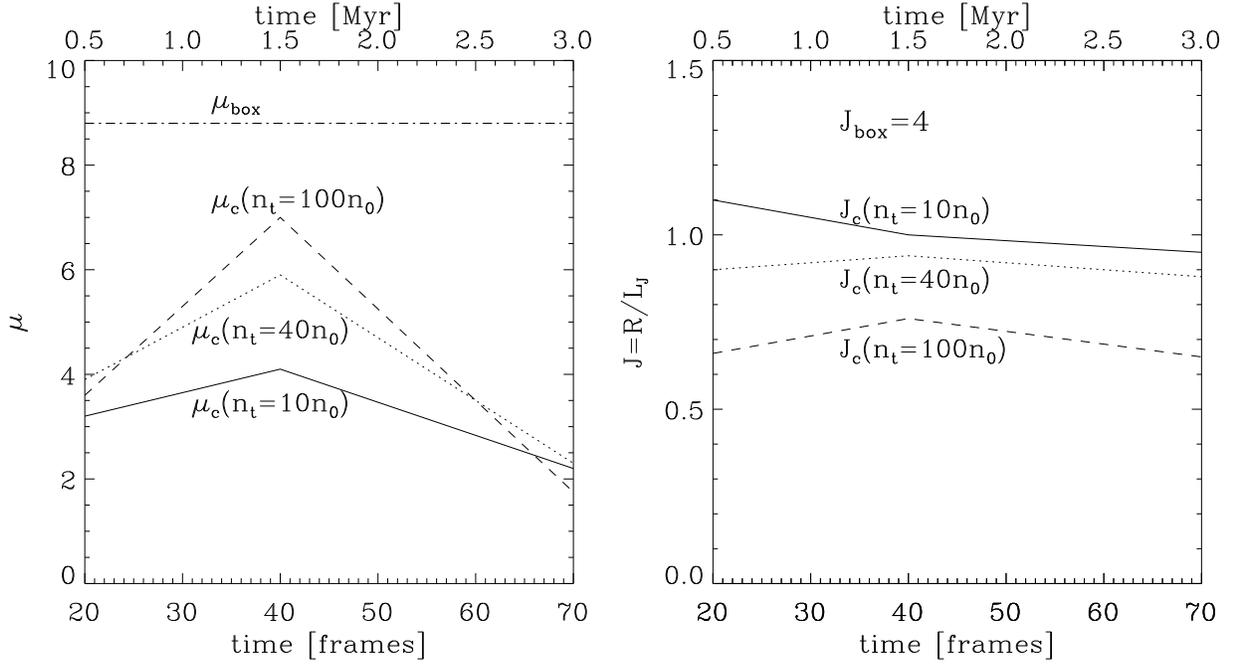}
\caption{Evolution of: ({\it left panel}) the mass-to-flux ratio (in
units of the critical value) $\mu_{\rm c}$ and of ({\it right panel})
the Jeans number $J_{\rm c}$ for the first long-lived rebounding clump
in the strongly supercritical run M10J4$\beta$1. The curve coding is as
in fig.\ \ref{fig:mu_J_mod_sup}. Note that this clump is always supercritical,
but almost always sub-Jeans. The failure to collapse even though $J_{\rm
c}$ is slightly above unity for $\nt=10 n_0$ at frame 20 can be
understood in terms of the non-accounting for turbulent support, and
also slight imprecisions in our estimates of the core parameters (see
text).}
\label{fig:mu_J_str_sup}
\end{figure}

These clumps last a time about
twice the longest durations seen in 
the subcritical case, and moreover have very well-defined identities,
contrary to the situation in the subcritical case. 
It is thus interesting
to investigate their physical properties. Figure
\ref{fig:mu_J_str_sup} shows the evoulution of $\mu_{\rm c}$ and
$J_{\rm c}$ for the first of these clumps from frame 20 to frame 70
($t=0.5$ to $t=3.$ Myr). It can be
seen that {\it this core is supercritical} ($\mu >1$), {\it but
sub-Jeans} ($J_{\rm c}< 1$) at all times and values of the
core-defining threshold density $\nt$, except at frame 20 and $\nt=10
n_0$, at which $J_{\rm c}$ is marginally larger than 1. Given the
approximate nature of our calculations (recall the core's radius is
defined assuming a spherical geometry), and that turbulent support is
not included in our analysis, we consider that the value $J_{\rm c} =
1.06$ found at that particular time and density threshold is still
consistent with the core not undergoing collapse in practice. We
conclude that these somewhat longer-lived, non-collapsing cores come
very close to collapse, but miss it, explaining their longer lifetimes
in terms of our estimate of \S \ref{sec:re-exp}. It appears unlikely
that the inclusion of AD in the calculations could help these cores to
collapse, as they are already supercritical, and so their support is
mainly thermal.

It is also interesting that these longest-lived cores arise in the
most strongly supercritical simulation, in which the magnetic support
is weakest. A possible explanation is that, somewhat
counterintuitively, turbulent flows with weak magnetic fields tend to
develop more moderate density fluctuations than either
non-magnetic or strongly magnetized ones, as has been
observed by various groups (Passot et al.\ 1995; Ostriker et al.\
1999; Heitsch et al.\ 2001; \BP\ \& Mac Low 2002). Thus, the production of
moderate-amplitude density fluctuations that can ``land'' close to
equilibrium without directly overshooting to collapse may be more likely
in the weak-field case.  

\subsubsection{Non-magnetic simulation} \label{sec:hd_run}

\begin{figure}
\plotone{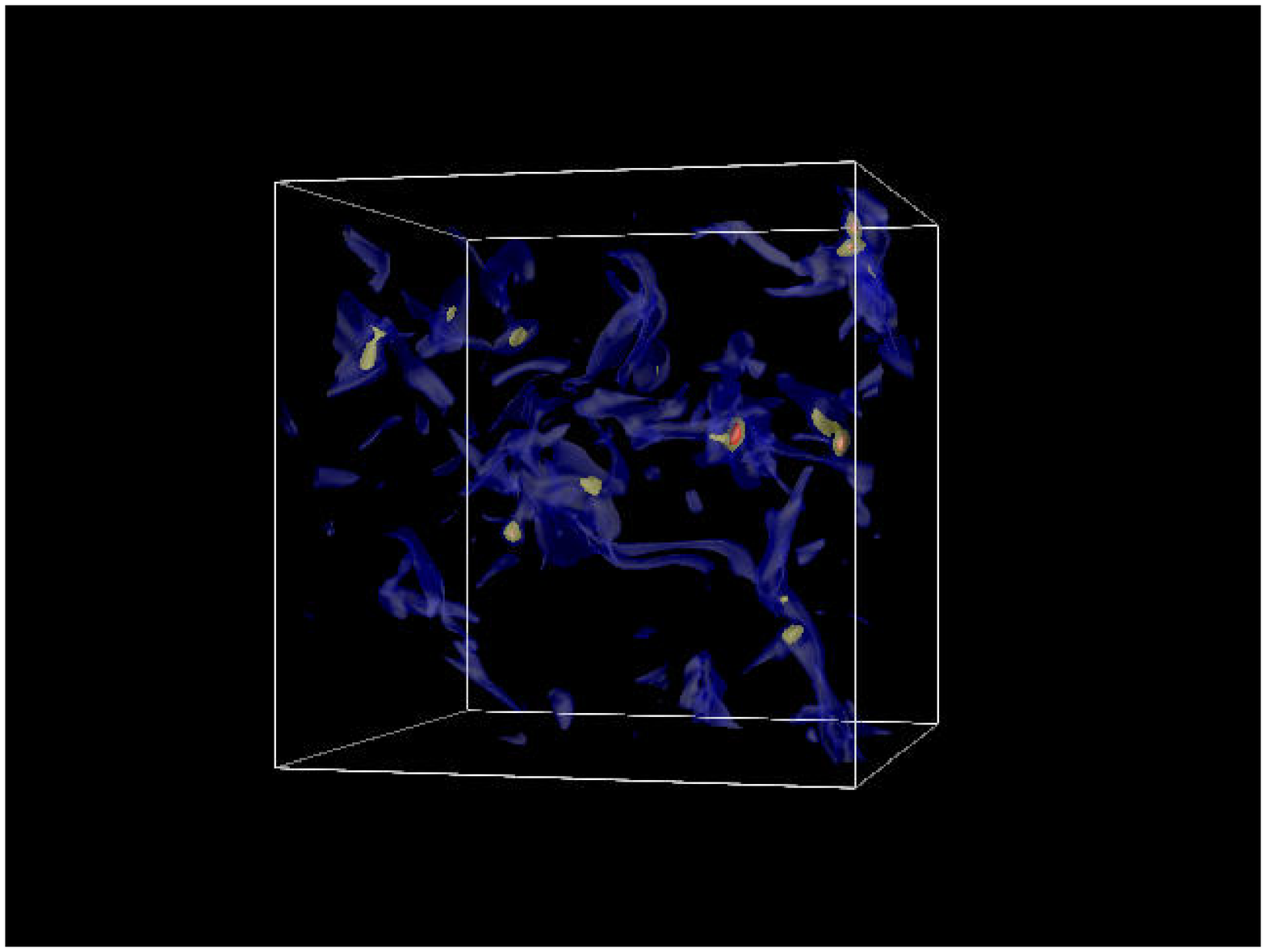}
\caption{Animation of the non-magnetic simulation
M10J4$\beta$$\infty$. The still frame in the paper edition shows the
final frame (\#44) of the animation. Note the larger number of cores in
comparison to either supercritical run.}
\label{anim:hd_run}
\end{figure}

Figure \ref{anim:hd_run} shows an animation of the nonmagnetic run
M10J4$\beta$$\infty$. This simulation stops after only 45 frames,
indicating that the collapse is more violent than in the magnetic
cases. We can only give lower bounds to the clump lifetimes in this
case, and a more detailed description must await a more robust handling
scheme for the collapsed objects, to be presented elsewhere.

Two features are worth noting here. First, {\it a
significantly larger number of collapsed objects and clumps} have
formed by the end of the run than at the same time in either MHD run.
This result must be considered preliminary only, as it is necessary to
verify that the larger number of cores is a real effect, and not just
a statistical fluctuation. For example, Heitsch et al.\ (2001)
reported that the statistical fluctuations between realizations with
identical parameters but different random initial conditions and
driving were larger than the systematic differences between magnetic
and non-magnetic cases. However, they did not discuss the number of
cores formed in their simulations. The difference in core number
between the magnetic and non-magnetic runs in our simulations is,
however, striking, and we plan to investigate this in a forthcoming paper
considering mean trends with the global parameters for ensembles of
simulations. 

The second point to note is that the time scales from formation to
full collapse seem to cover a wider range, of $\sim 1/2$ to 1
Myr, with the fastest-collapsing objects being those that involve
collisions of clumps. Moreover, a population of 
non-collapsed objects is seen by the end of the run for which we
cannot observe the final fate nor the corresponding time scale. 
Several re-expanding clumps are also observed. The investigation of these
trends must also await a more robust handling scheme for the sink cells.

Nevertheless, these preliminary results lead us to speculate that the
presence of 
a magnetic field can further reduce the efficiency of collapsed-object
formation in comparison with that of an equivalent non-magnetic case
mostly by reducing the probability of core formation, rather than by
significantly delaying the collapse of individual objects.

\section{Discussion, implications and caveats}
\label{sec:discussion}

Several interesting points can be noted about the results described above:

1. Our results suggest that, at least under the assumptions of our
numerical simulations, molecular cloud clumps and
cores are in general likely to be dynamic, out-of-equilibrium
structures, rather than quasi-hydro/magneto-static structures. This is
however not in conflict with observations, as several
studies show that synthetic observations of the clouds and the clumps
within them in numerical simulations compare favorably to observations
(e.g., Padoan et al.\ 1999, 2003; Klessen 2001; Ostriker et al.\ 2001;
\BP\ \& Mac Low 2002; Gammie et al.\ 2003; Klessen, \BP, \VS\ \& Dur\'an
2004; Schmeja \& Klessen 2004). In
particular, dynamically-formed cores can have angle-averaged column
density profiles that resemble that of a BE-sphere (\BP, Klessen \& \VS\
2003), or in fact various other functional forms (Harvey et al.\
2003). This is because much detail is lost by the line-of-sight and
angle averaging.

2. The clump/core lifetimes that we have found in this paper (1/2-1 Myr
for collapsing objects, 1/2-3 Myr for non-collapsing ones) also
compare favorably with observational estimates of core lifetimes, such
as those of Lee \& Myers (1999), who obtained estimated lifetimes of
0.3-1.6 Myr. However, an interesting remark is in order.
Lee \& Myers obtained those estimates under the assumption that all
cores undergo AD-mediated contraction, infall, and then protostellar
emission, so that the number of starless cores ($n_{\rm SL}$) and of
cores with young stellar objects (YSOs) ($n_{\rm YSO}$) are
respectively proportional to the time spent 
in the prestellar and protostellar epochs during this
process. However, our results advocate the possibility that not all
clumps undergo 
collapse, and in this case it is easy to show that the ratio $n_{\rm
SL}/n_{\rm YSO}$ is an upper bound to the lifetime ratio. Indeed,
define $e \equiv n_{\rm SL}/n_{\rm YSO}$. If the number of starless
cores comprises a fraction $\epsilon$ of failed cores and a fraction
$1-\epsilon$ of true pre-stellar cores (with number $n_{\rm PS}$), then
\begin{equation}
e \equiv \frac{n_{\rm SL}} {n_{\rm YSO}} = \frac{n_{\rm PS}} { n_{\rm
YSO}}\left(\frac{1}{1 - \epsilon}\right) .
\end{equation}
Now, since the ratio of true pre-stellar objects to proto-stellar ones
is indeed equal to the ratio of pre-stellar to protostellar lifetimes
$\tau_{\rm PS}/\tau_{\rm YSO}$, we find
\begin{equation}
e = \frac{\tau_{\rm PS}}{\tau_{\rm YSO}} \left(\frac{1}{1 - \epsilon}\right) >
\frac{\tau_{\rm PS}}{\tau_{\rm YSO}},
\label{eq:PS-YSO_ratio}
\end{equation}
which is the desired result. Although the actual fraction of failed
cores has not been quantified in our numerical simulations (again,
ensembles of simulations are needed), and is unknown in the
observations, this consideration is consistent with the fact that
the longest lifetimes of {\it collapsing} objects in our simulations are
shorter by a factor of $\sim 2/3$ than the upper bound of Lee \& Myers'
(1999) estimates.

3. It is important to compare the fraction of collapsing versus
non-colapsing clumps in the simulations with observational data. This
will be attempted in detail in a separate paper, but here we can give
the following simple estimate. Assuming that a minimum number density of
$\sim 10^4$ cm$^{-3}$ is necessary to excite a high-density tracer such
as ammonia, a typical ``ammonia core'' consists of the ``tip of the
iceberg'' region of a clump above this minimum density. Such minimum
density can be identified with our core-defining threshold density 
$\nt$, giving $\nt \sim 20 n_0$. In our simulations, typical failed,
re-expanding cores defined with this value of $\nt$ only reach mean
densities of up to 2--3 times $\nt$ ($\sim 50 n_0 = 2.5 \times 10^4$
cm$^{-3}$). Thus, it seems that most of the re-expanding clumps and
cores in our 
simulations would lie below the ammonia detection limit, and would
preferentially appear in lower-density-tracer surveys, while surveys
in high-density tracers such as ammonia may tend to preferentially select
collapsing objects. Detailed synthetic observations of the simulations
with various clump-defining density thresholds are necessary to
determine whether the observed fraction of inward versus outwards
(i.e., contracting vs.\ re-expanding) motions in cores (e.g., Lee, Myers
\& Tafalla 2001) are well reproduced by the simulations.

4. In the analysis of \S \ref{sec:core_props}, we have
neglected turbulent support of the clumps and cores. Interestingly,
the ($J_{\rm c},\mu_{\rm c}$) parameter space seems to describe very
well the outcome of the clump/core system evolution. This may be
because at the small sizes of these objects, the internal turbulence
may be of secondary importance. Indeed, real cores are typically
transonic or even subsonic, a feature reproduced by the simulations
(Klessen et al.\ 2004). In this case, the turbulent support is
comparable or smaller than the thermal support, and so only the magnetic
and thermal supports remain for consideration. 

5. The caveat that our results hold only within the realm of the
assumptions described in \S \ref{sec:assumptions} should be
stressed. Although in that section we have referred to observational
evidence suggestive that the numerical setup and parameters used in
this paper are the most realistic (except for the neglect of AD), the
issue of whether molecular clouds are nearly isothermal, continually
driven and supercritical is still unsettled. Our clump lifetimes in the
subcritical case are short because the turbulence is continually
driven. We cannot overemphasize
the importance of continued observational surveys directed at
determining the true conditions in molecular clouds, or the fraction of
clouds under each type of condition. Also, numerical simulations of the
interstellar medium at large, in which the formation of molecular
clouds is accurately simulated, can shed light on this issue.

\section{Summary and conclusions} \label{sec:conclusions}

In this paper we have discussed the lifetimes of clumps and cores in both the
non-magnetic and magnetic cases, in single-phase, isothermal,
turbulent models of molecular clouds. Under these conditions, we have
remarked that, in the non-magnetic case, the hotter, low-density gas
necessary for the confinement of {\it stable} BE-type 
hydrostatic configurations is absent, and so we do not expect these
structures to form out of transient turbulent fluctuations in the
clouds. We have then given a qualitative discussion 
suggesting that non-truncated (``extended'') isothermal structures 
may be expected to be unstable in
general, in agreement with the fact that non-magnetic
polytropic simulations with $\gamef < 4/3$ do not form hydrostatic
structures (e.g., \VS\ et al. 1996; Klessen et al.\ 2000; Bate et al.\
2003; Li et al.\ 2003). Thus, we have suggested that in the non-magnetic case, 
clumps should be expected to be completely transient structures, either
proceeding to collapse right away upon the turbulent compression that
formed them, or else re-expanding to merge back into their
environment (``failed'' clumps). 

For failed clumps, we have given a simple estimate for the characteristic
re-expansion time, as slowed down by the self-gravity of the clump,
finding times $\sim$ a few clump free-fall times, comparable to the
lifetime estimates of Lee \& Myers (1999). The longest lifetimes
correspond to re-expanding clumps that come very close to equilibrium
between self-gravity and thermal pressure. In general, failed clumps are not
destined to form stars. However, in the simulations there is always
a population of these clumps present, in agreement with some
observational suggestions (Taylor et al.\ 1996). These can be
associated with at least part of the observed starless cores, which
traditionally have been thought of as being in a pre-stellar phase.

In the magnetic case, we have presented a set of of driven-turbulence,
ideal-MHD numerical simulations focusing on the lifetimes and
criticality of the individual clumps and cores in relation to the
conditions in their parent clouds. We have found that: 

a) Magnetostatic 
clumps {\it do not} form in the case of subcritical environments; the
clumps that form in this case have mean densities $\nm \sim$
10--100$n_0$ ($5\times 10^3$--$5\times 10^4$ cm$^{-3}$), and 
lifetimes $\sim$ 0.2--1.3 Myr, ultimately being dispersed by the
turbulence in the environment. A few of these clumps nevertheless
become moderately gravitationally bound, being the ones with the longest
lifetimes. An estimate of the AD time scale in these clumps gives
characteristic times $\gtrsim 1.3$ Myr, suggesting that they can 
marginally be ``captured'' by AD and increase
their mass-to-flux ratio before they can be dispersed, thus increasing
their gravitational binding, to eventually become supercritical and
collapse. Thus, they adjust to the standard model scenario, except
for the fact that with these time scales, AD does not significantly
increase the cores' lifetimes 
in comparison to those of supercritical or non-magnetic ones. This
questions the usefulness of AD-mediated contraction as a
SFE-decreasing agent in the standard model, and the
criticality of the parent may become irrelevant for the formation of
supercritical cores in short time scales.

b) In the supercritical case, dense cores ($\nm \sim$ 100--1000 $\times
n_0$, or $5\times 10^5$--$5\times 10^6$ cm$^{-3}$
do form, being supercritical since their early stages
and collapsing on time scales $\sim 2 \tfc \sim 1$ Myr. Ambipolar
diffusion is not required for collapse in this case, 
as the objects rapidly become supercritical by the turbulent
compressions themselves, since there is a large enough mass reservoir
around them. The 
lifetimes in the magnetic case are in general again in 
agreement with the observational estimates of Lee \& Myers (1999) and
our own simple, non-magnetic estimate.

c) As in the non-magnetic case, coexisting with the collapsing cores,
numerous lower-density ($\nm \lesssim 100 n_0$) ``failed'' ones
form and disperse in times comparable to their free-fall times.
These ``failed'' clumps include the longest-lived objects observed in our
simulations, again in agreement with our simple calculation, which
suggests that the longest lifetimes correspond to 
re-expansion from states very close to equilibrium. However, it appears
unlikely that the inclusion of ambipolar
diffusion could induce their collapse, as these objects are already
supercritical, and the reason they do 
not collapse is because they do not manage to become gravitationally
unstable with respect to the thermal support.

Within the framework of our assumptions, our results support the
developing notion that star formation is a 
dynamic process, in which the only hydrostatic configurations 
may be the stars themselves.

\acknowledgements

We have benefitted from comments and constructive criticisms by
Shantanu Basu, Edith Falgarone, Daniele Galli, Lee Hartmann, Ralf
Klessen, Susana Lizano, Mordecai Mac Low, Lee Mundy, John Scalo and
Mario Tafalla.  
E.V.-S. thankfully acknowledges the hospitality of
the KAO (Korea Astronomy Observatory) and Chungnam National University, Korea. 
We acknowledge partial financial support from CONACYT grants
27752-E and 36571-E to E.V.-S and I 39318-E to J. B.-P., and from Ferdowsi
University to M.S. J.K. was supported by the Astrophysical Research
Center for the Structure and Evolution of the Cosmos (ARCSEC) of the
Korea Science and Engineering Foundation 
through the Science Research Center (SRC) program. The numerical simulations
were performed on the Linux cluster at KAO,
with funding from KAO and ARCSEC. This work has
made extensive use of NASA's Astrophysics Abstract Data Service and
LANL's astro-ph archives.

\clearpage

\begin{deluxetable}{cccccccccccc}
\tablecaption{Numerical simulation parameters.\tablenotemark{a}
\label{tab:runs}} 
\tablewidth{0pt}
\tablehead{
\colhead{Run name} & \colhead{$\Ms$}   & \colhead{$J$}   &
\colhead{$\beta$} & Resolution &
\colhead{$L$(pc)}  & \colhead{$n_0$(cm$^{-3}$)} & \colhead{$B_0$($\mu$G)} &
\colhead{$\sigma$(km s$^{-1}$)}\tablenotemark{b}     & 
\colhead{$\mu$} 
}
\startdata
M10J4$\beta$$\infty$	&10& 4	&$\infty$&$256^3$ & 4	& 500	& 0.0	& 2.0	& $\infty$\\
M10J4$\beta$1	& 10	& 4	& 1	& $256^3$ & 4	& 500	& 4.6	& 2.0	& 8.8\\
M10J4$\beta$.1& 10	& 4	& 0.1	& $256^3$ & 4	& 500	& 14.5	& 2.0	& 2.8 \\
M10J4$\beta$.01& 10	& 4	& 0.01	& $256^3$ & 4	& 500	& 45.8	& 2.0	& 0.9\\
\\
M10J4$\beta$.01LR& 10	& 4	& 0.01	& $128^3$ & 4	& 500	& 45.8	& 2.0	& 0.9\\
M10J3$\beta$.01& 10	& 3	& 0.01	& $128^3$ & 4	& 281	& 34.4	& 2.0	& 0.7\\
\\
M8J4$\beta$.1& 8	& 4	& 0.1	& $128^3$ & 2.56	& 1220	& 22.6	& 1.6	& 2.8 \\
M8J4$\beta$.01& 8	& 4	& 0.01	& $128^3$ & 2.56	& 1220	& 71.6	& 1.6	& 0.9 \\

%

 \enddata




\tablenotetext{a} {See description in \S \ref{sec:num_meth}.}

\tablenotetext{b} {Turbulent velocity dispersion in the computational box.}


\end{deluxetable}

\clearpage

\begin{figure}
\plotone{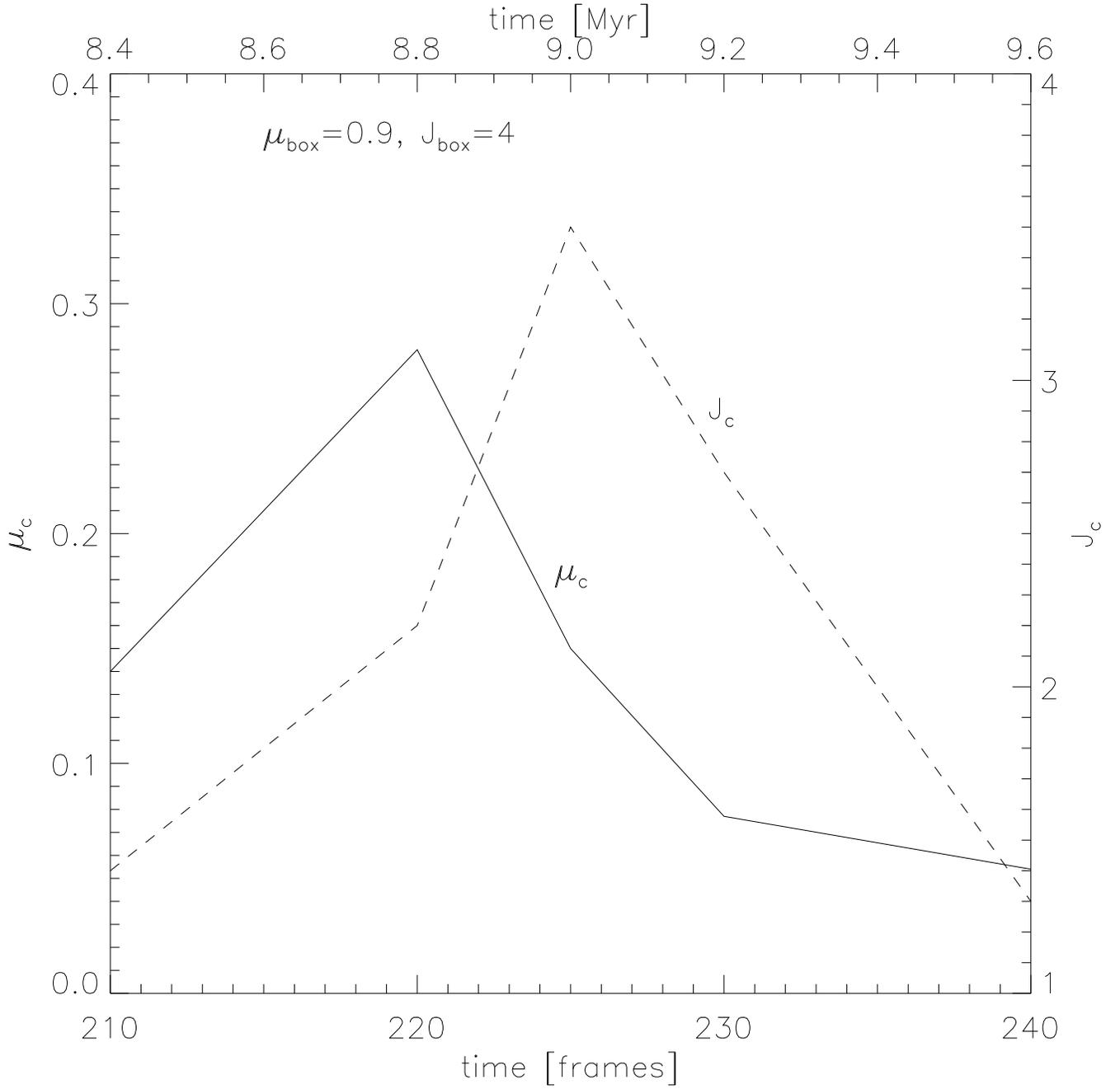}
\caption{Evolution of the mass-to-flux ratio (in units of the critical
value) $\mu_{\rm c}$ and of the Jeans number $J_{\rm c}$ for the
densest, moderately long-lived core in the subcritical run
M10J4$\beta$.01. The lower time-axis is given in
units of the frame number in the animation.}
\label{fig:mu_J_sub}
\end{figure}

\appendix
\section{Ambipolar diffusion time scale in a subcritical clump}
\label{app:AD_time} 

Here we estimate the ambipolar diffusion (AD) time scale $\tad$ under
the conditions of the longest-lived clump in the subcritical run
M10J4$\beta$.01. To this end, we use the formulas given by Fiedler \&
Mouschovias (1992). They write

\begin{equation}
\tau_{\rm AD}=\frac{8}{1.4\pi^2} \frac{\tau_{\rm ff}^2}{\tau_{\rm
ni }} \left(1-\frac{8}{\pi^2}\frac{\tau_{\rm ff}^2}{\tau_{\rm
s}^2}\right)^{-1}, 
\label{eq:t_ad}
\end{equation}
where $\tau_{\rm s}\equiv R_{\rm c}/c_{\rm s}$ is the sound
crossing time and $\tau_{\rm ff}$ is free-fall time scale
\begin{equation}
\tau_{\rm ff}=\left(\frac{3\pi}{32G\rho_{\rm n}}\right)^{1/2}.
\end{equation}
Note that this definition of the free-fall time is smaller by a factor of
$\sqrt{3/32}$ than that used in the main body of this paper. This,
however, causes no inconsistency, as we compare the AD time scale and
core lifetimes in physical units. Finally,
$\tau_{\rm ni}$ is the neutral-ion collision time scale, given by
\begin{equation}
\tau_{\rm ni}=\frac{m_{\rm i}+m_{\rm n}}{\rho_{\rm
i} \langle\sigma w\rangle_{\rm in}},
\end{equation}
in which $m_{\rm n}$ and $m_{\rm i}$ are the neutral and ion particle
masses, respectively, $\rho_{\rm n}$ and $\rho_{\rm i}$ the
neutral and ion mass densities, and $\langle\sigma w\rangle_{\rm in}$ is the
neutral-ion collision rate. We can write the above equation as
\begin{equation}
\tau_{\rm ni}=\frac{1+\frac{m_{\rm n}}{m_{\rm
i}}}{n_{\rm i}\langle\sigma w\rangle_{\rm in}}.
\end{equation}
Fiedler \& Mouschovias also give an approximate expression for the ion
number density as
\begin{equation}
n_{\rm i} \approx K \left(\frac{n_{\rm n}}{10^5 {\rm cm}^{-3}}\right)^k,
\end{equation}
with canonical values $K=3\times 10^{-3}$ cm$^{-3}$,
$k=1/2$, and $\langle\sigma w\rangle_{\rm in}=1.69\times 10^{-9} {\rm
cm}^{3} {\rm s}^{-1}$. We also take $m_{\rm i} \sim 29.0 m_{\rm H}$ and
$m_{\rm n} = 2.36 m_{\rm H}$, with $m_{\rm H} = 1.67 \times 10^{-24}$ g. 

Defined out to a threshold density $\nt = 40 n_0$, the core in question
has a mean density of $55 n_0 = 2.75 \times 10^4$ cm$^{-3}$ at the
time when it attains the maximum value of its mass-to-flux ratio,
$\mu_{\rm c,max} = 0.28$ (see fig.\ \ref{fig:mu_J_sub}). This core is
highly flattened, and thus we compute its radius as the maximum distance
between the density maximum and the clump's periphery (cf. \S
\ref{sec:core_analysis}), finding $R_{\rm c} 
= 0.3$ pc. With these values we thus obtain $\tau_{\rm ni}= 1.29
\times 10^4$ yr, $\tau_{\rm ff} = 2.02 \times 10^5$ yr,
$\tau_{\rm s} = 1.5 \times 10^6$ yr, and $\left[1-(8/\pi^2)(\tau_{\rm
ff}/\tau_{\rm s})^2\right]^{-1}=1.015$.
Substituting in eq.\ (\ref{eq:t_ad}) we finally obtain
\begin{equation}
\tau_{\rm AD} \approx 9.2 \tau_{\rm ff} = 1.9~{\rm Myr}.
\end{equation}

This estimate is very close to the simpler, more general estimate often
quoted (e.g., Ciolek \& Basu 2001) of $\tau_{\rm AD} \approx \tau_{\rm
ff}^2/ (1.4 \tau_{\rm ni}) = 11.2 \tau_{\rm ff} = 2.3$ Myr.

A correction to this estimate to take into account the initial value of the
mass-to-flux ratio in the core, $\mu_{\rm c}$ has recently been given by
Ciolek \& Basu (2001). From their fig.\ 1 and the value $k=1/2$ that we
have used here, 
the correction factor is $C(\mu_{\rm c0}) \sim 0.7$. This is the
smallest possible value of their correction factor at this value of
$\mu_{\rm c}$. So we conclude that 
\begin{equation}
\tau_{\rm AD} \gtrsim 1.3~{\rm Myr}.
\end{equation}


\begin{references}


\reference{} Alves, J. F., Lada, C. J., \& Lada, E. A. 2001, Nature 409, 159



\reference{} Ballesteros-Paredes, J. 2004, in From Observations to
Self-Consistent Modeling of the Interstellar Medium, eds. M. de
Avillez and D. Breitschwerdt (Dordrecht: Kluwer), 67

\reference{} Ballesteros-Paredes, J., \& V\'azquez-Semadeni, E.
1997, in ``Star Formation, Near and Far. 7th Annual Astrophysics
Conference in Maryland'', ed. S. Holt and L. Mundy (New York: AIP
Press ), p.81

\reference{1} Ballesteros-Paredes, J., V\'azquez-Semadeni, E., \&
Scalo, J. 1999a, ApJ, 515, 286

\reference{}  Ballesteros-Paredes, J., Hartmann, L. \&
V\'azquez-Semadeni, E. 1999b, ApJ 527, 285 

\reference{} Ballesteros-Paredes, J.~\& Mac Low, M.\ 2002, \apj, 570, 734 

\reference{} Ballesteros-Paredes, J., Klessen, R. \& \VS, E. 2003, ApJ
592, 188

\reference{} Bania, T.M. \& Lyon, J.G. 1980, ApJ 239, 173


\reference{} Bate, M. R., Bonnell, I. A., \& Bromm, V. 2002, MNRAS 336, 705


\reference{} Blitz, L., 1993, in ``Protostars and Planets III'',
eds. E. H. Levy and J. I. Lunine (Tucson: Univ. of Arizona Press), 125

\reference{} Blitz, L., \& Williams, J P. 1999, in ``The Origin of Stars
and Planetary Systems'', ed. C. J. Lada and N. D. Kylafis (Dordrecht:
Kluwer), 3 

\reference{} Bonnor, W. B. 1956, MNRAS, 116, 351

\reference{} Bourke, T. L., Myers, P. C., Robinson, G., Hyland,
A. R. 2001, ApJ 554, 916

\reference{} Brunt, C. M., \& Heyer, M. H. 2002, ApJ 566, 289

\reference{} Brunt, C. M. 2003, ApJ 583, 280

\reference{} Ciolek, G. E. \& Basu, S. 2001, ApJ 547, 272

\reference{} Clark, P. C. \& Bonnell, I. A. 2004, MNRAS, in press
(astro-ph/0311286)

\reference{} Clarke, C. J. \& Pringle, J. E. 1997, MNRAS 288, 674

\reference{} Chandrasekhar, S. 1961, {\it Hydrodynamic and Hydromagnetic
Stability} (New York: Dover)

\reference{} Chandrasekhar, S., \& Fermi, E. 1953, ApJ, 118, 116

\reference{} Crutcher, R. M. 1999, ApJ 520, 706

\reference{} Crutcher, R. 2004, in ``Magnetic Fields and
Star Formation: Theory versus Observations", eds. Ana I. Gomez de
Castro et al, (Dordrecht: Kluwer Academic Press), in press

\reference{} Crutcher, R., Heiles, C. \& Troland, T. 2002, in Simulations
of Magnetohydrodynamic Turbulence in Astrophysics, ed. T. Passot \&
E. Falgarone (Berlin: Springer), 155

\reference{3} Curry, C. L. 2000, ApJ, 541, 831





\reference{} Ebert, R.\ 1955, Zeitschrift f{\"u}r Astrophysik, 36, 222



\reference{4} Elmegreen, B. G. 1993, ApJ, 419, L29


\reference{} Evans, N. J., II 1999, ARAA 37, 311

\reference{} Evans, N.~J., Rawlings, J.~M.~C., Shirley, Y.~L., \&
Mundy, L.~G.\ 2001, \apj, 557, 193 



\reference{} Fiedler, R. A. \& Mouschovias, T. Ch. 1992, ApJ 391, 199



\reference{} Field, G. B., Goldsmith, D. W., \& Habing, H. J. 1969, ApJ,155, L149


\reference{} Gammie, C. F., Lin, Y.-T., Stone, J. M., \& Ostriker,
E. C. 2003, ApJ 592, 203

\reference{} Gazol, A., Passot, T. \& Sulem, P. L. 1999, Phys. Plasmas
6, 3114


\reference{} Gilden, D. L. 1984, ApJ 283, 679

\reference{} Goldsmith, P. F. 1987, in Interstellar Processes, eds./
D. J. Hollenbach and H. A. Thronson (Dordrecht: Reidel), 51




\reference{} Hartmann, L., Ballesteros-Paredes, J., \& Bergin, E.
A. 2001, ApJ, 562, 852

\reference{}  Harvey, D.~W.~A., Wilner, D.~J., Lada, C.~J., Myers,
P.~C., Alves, J., \& Chen, H.\ 2001, \apj, 563, 903

\reference{}  Harvey, D.~W.~A., Wilner, C.~J., Myers, P.~C. \& Tafalla,
M.~2003, ApJ 597, 424 


\reference{8} Hayashi, C. 1966, AR\&A 4, 171


\reference{} Heitsch, F., Mac Low, M. M., \& Klessen, R. S. 2001,
ApJ, 547, 280



\reference{} Heyer, M. H. \& Brunt, C. M. 2004, ApJ, submitted

\reference{} Hunter, C. 1977, ApJ 218, 834


\reference{} Hunter, J. H., Jr., Sandford, M. T., II, Whitaker, R. W.,
Klein, R. I. 1986, ApJ 305, 309

\reference{} Jijina, J., Myers, P. C., \& Adams, F. C. 1999, ApJS 125, 161

\reference{} Johnstone, D., Wilson, C.~D., Moriarty-Schieven, G.,
Joncas, G., Smith, G., Gregersen, E., \& Fich, M.\ 2000, \apj, 545, 327





\reference{} Kim, J., Ryu, D., Jones, T. W. \& Hong, S. S. 1999, ApJ
514, 506

\reference{} Kim, J., Balsara, D. \& Mac Low, M.-M. 2001, JKAS 34, S333

\reference{} Klessen, R. S., Heitsch, F., \& MacLow, M. M. 2000,
ApJ, 535, 887

\reference{} Klessen, R. S. 2001, ApJ 556, 837

\reference{} Klessen, R. S., \BP, J., \VS, E. C. \& Dur\'an, C. 2004,
ApJ, submitted (astro-ph/0306055) 



\reference{} Koyama, H. \& Inutsuka, S.-I. 2002, ApJ 564, L97

\reference{} Larson, R. B. 1969, MNRAS, 145, 271

\reference{} Larson, R. B. 1981, MNRAS, 194, 809

\reference{} Larson, R. B. 1985, MNRAS, 214, 379

\reference{} Lee, C. W. \& Myers, P. C. 1999, ApJS 123, 233

\reference{} Lee, C. W., Myers, P. C. \& Tafalla, M. 2001, ApJS 136, 703

\reference{} Li, P. S., Norman, M. L., Mac Low, M.-M. \& Heitsch,
F. 2004, ApJ 605, 800

\reference{} Li, Y., Kleesen, R. S. \& Mac Low, M.-M. 2003, ApJ 592, 975

\reference{} Li, Z.-Y. \& Nakamura, F. 2004, ApJL, in press (astro-ph/0405615)

\reference{} Lizano, S., \& Shu, F. H. 1989, ApJ, 342, 834

\reference{} Lombardi, M.~\&  Bertin, G.\ 2001, \aap, 375, 1091

\reference{} Low, C. \& Lynden-Bell, D. 1976, MNRAS 176, 367

\reference{} Mac Low, M.-M. \& Klessen, R. S. 2004, Rev. Mod. Phys. 76, 125

\reference{} Maddalena, R. J. \& Thaddeus, P. 1985, ApJ 294, 231

\reference{} Masunaga, H. \& Inutsuka, S.-I. 1999, ApJ 510, 822


\reference{} McKee, C. F., Zweibel, E. G., Goodman, A. A. \& Heiles,
C. 1993, in Protostars and Plantes III, eds. E.H. Levy and J.I. Lunine
(Tucson: Univ. of Arizona Press), p. 327


\reference{} McLaughlin, D. E., \& Pudritz, R. E. 1996, ApJ, 469, 194

\reference{} Mestel, L., Spitzer, L., Jr. 1956, MNRAS 116, 503

\reference{} Mouschovias, T. C.  1976a, ApJ 207, 141

\reference{} Mouschovias, T. C.  1976b, ApJ 206, 753

\reference{} Mouschovias, T. C. \& Spitzer, L., Jr. 1976, ApJ 210, 326

\reference{} Myers, P. C. 1978, ApJ 225, 380







\reference{} Nakano, T. \& Nakamura, T. 1978, PASJ 30, 671

\reference{} Nakano, T. 1998, ApJ 494, 587

\reference{} Ossenkopf, V. \& Mac Low, M.-M. 2002, A\&A 390, 307

\reference{} Ostriker, J. 1964, ApJ 140, 1056

\reference{} Ostriker, J. 1965, ApJS 11, 167

\reference{} Ostriker, E. C., Gammie, C. F. \& Stone, J. M. 1999, ApJ 513, 259

\reference{} Ostriker, E. C., Stone, J. M. \& Gammie, C. F. 2001, ApJ 546, 980

\reference{} Padoan, P. 1995, MNRAS, 277, 377

\reference{} Padoan, P. \& Nordlund, \AA. 1999, ApJ 526, 279

\reference{} Padoan, P., Bally, J., Billawala, Y., Juvela, M.,
Nordlund, \AA\ 1999, ApJ 525, 318

\reference{} Padoan, P., Juvela, M., Goodman, A.A., Nordlund, \AA 2001
ApJ 553, 227

\reference{} Padoan, P., Goodman, A. A. \& Juvela, M. 2003, ApJ 588, 881



\reference{} Passot, T., \VS, E. \& Pouquet, A. 1995, ApJ 455, 536


\reference{} Passot, T. \& \VS, E. 2003, A\&A 398, 845

\reference{} Pouquet, A., L\'eorat, J. \& Passot, T. 1991, in Advance in
Turbulence III, eds. A. Johansson and H. Alfredsson (Berlin:Springer), 343

\reference{} Pratap, P., Dickens, J. E., Snell, R. L.,  Miralles, M. P.,
Bergin, E. A., Irvine, W. M., \& Schloerb, F. P. 1997, ApJ 486, 862


\reference{}  Scalo, J. 1987, in Interstellar Processes,
eds. D.J. Hollenbach and H.A. Thronson (Dordrecht: Reidel), p. 349


\reference{} Scalo, J., V\'azquez-Semadeni, E., Chappell, D., \&
Passot, T. 1998, ApJ 504, 835

\reference{} Schmeja, S. \& Klessen, R. S. 2004, A\&A 419, 405

\reference{11} Shadmehri, M., V\'azquez-Semadeni, E., \&
Ballesteros-Paredes, J. 2002, in Seeing Through the Dust: The
Detection of HI and the Exploration of the ISM in Galaxies,
eds. A. R. Taylor, T. L. Landecker, and A. G. Willis (San Francisco:
Astronomical Society of the Pacific), 190




\reference{} Shu, F. 1977, ApJ 214, 488

\reference{14} Shu, F. H., Adams, F. C., \& Lizano, S. 1987, ARA\&A, 25, 23




\reference{} Spaans, M. \& Silk, J. 2000, ApJ 538, 115

\reference{} Spitzer, L. 1968. Diffuse Matter in Space (New York,
Wiley), \S 6.1

\reference{} Stone, J. M., Ostriker, E. C., \& Gammie, C. F. 1998,
ApJ, 508, L99 

\reference{} Taylor, S. D., Morata, O., \& Williams, D. A. 1996, A\&A
313, 269

\reference{} Tilley, D. A. \& Pudritz, R. E. 2004, MNRAS, submitted
(astro-ph/0406122) 




\reference{} Truelove, J. K., Klein, R. I., McKee, C. F., Hilliman,
J. H. II., Howell, L. H., \& Greenough, J. A. 1997, ApJ, 489, L179

\reference{} V\'azquez-Semadeni, E., Passot, T., \& Pouquet, A.
1996, ApJ, 473, 881

\reference{} V\'azquez-Semadeni, E., Ostriker, E. C., Passot, T., Gammie, C.
\& Stone, J., 2000, in ``Protostars \& Planets IV'', ed. V. Mannings,
A. Boss \& S. Russell (Tucson: Univ.\ of Arizona Press), 3

\reference{} V\'azquez-Semadeni, E. 2002, in Seeing Through the Dust:
The Detection of HI and the Exploration of the ISM in Galaxies,
eds. R. Taylor, T. Landecker, \& A. Willis (ASP: San Francisco), 155

\reference{} V\'azquez-Semadeni, E., Ballesteros-Paredes, J. \& Klessen, 
R. 2003a, ApJ 585, L131

\reference{} V\'azquez-Semadeni, E., Ballesteros-Paredes, J. \& Klessen, 
R. 2003b, in Galactic Star Formation Across the Stellar Mass
Spectrum, ASP conf. series, ed. J. M. De Buizer (San Francisco: ASP), in 
press (astro-ph/0206038)

\reference{} V\'azquez-Semadeni, E. 2003, in Star Formation at High
Angular Resolution, IAU Symposium 
221, eds. M. Burton, R. Jayawardhana \& T. Bourke (San Francisco:
Astronomical Society of the Pacific), in press (astro-ph/0309717)


\reference{} Walder, R. \& Folini, D. 1998 A\&A 330, L21

\reference{} von Weizs\"acker, C.F. 1951, ApJ 114, 165

\reference{} Yabushita, S. 1968, MNRAS 140, 109

\reference{} Zuckerman, B. \& Evans, N. J. II 1974, ApJ 192, L149

\reference{} Zweibel, E. G. 1990, ApJ 348, 186


\end{references}
\end{document}